\documentclass[12pt]{article}
\usepackage{graphicx}
\def \lesssim{\stackrel{<}{\sim}}

\def \hf{\frac{1}{2}}

\def \bea{\begin{eqnarray}}
\def \beq{\begin{equation}}

\def \eea{\end{eqnarray}}
\def \eeq{\end{equation}}

\def \({\left(}
\def \){\right)}
\def \[{\left[}
\def \]{\right]}

\def \half{\frac{1}{2}}
\def \s{\sqrt{2}}

\def \ep{\epsilon}

\begin{document}
\rightline{TECHNION-PH-15-1}
\rightline{arXiv:1501.03272}
\vskip 10mm
\centerline{\bf SU(3) in $D$ decays: From $30\%$ symmetry breaking to $10^{-4}$ precision}
\bigskip
\centerline{Michael Gronau}
\medskip
\centerline{\it Physics Department, Technion -- Israel Institute of Technology}
\centerline{\it Haifa 32000, Israel}
\bigskip
\begin{quote}
Flavor SU(3) symmetry, including $30\%$ first order SU(3) breaking, has been shown 
to describe adequately a vast amount of data for charmed meson decays to two pseudoscalar
mesons and to a vector and a pseudoscalar meson. We review a recent dramatic progress 
achieved by applying a high order perturbation expansion in flavor SU(3) breaking and treating carefully isospin breaking.  
We identify a class of U-spin related $D^0$ decays to pairs involving charged pseudoscalar or 
vector mesons, for which high-precision nonlinear amplitude relations are predicted. Symmetry breaking terms affecting these relations are fourth order U-spin breaking, and terms 
which are first order in isospin breaking and second order in U-spin breaking.  
The predicted relations are shown to hold 
within experimental errors at a precision varying between $10^{-3}$ and $10^{-4}$, 
in agreement with estimates of high order terms. 
Improved branching ratio measurements for $D^0 \to K^+\rho^-, K^{*+}\pi^-$ and for 
decay modes involving three other pairs of charged pseudoscalar and vector mesons 
could further sharpen two of these precision tests. We also obtain
amplitude relations for $D^0$ decays to pairs of neutral pseudoscalar mesons, and relations for rate asymmetries between decays involving $K^0_S$ and $K^0_L$, which hold up to second order U-spin breaking at a level of several percent.
\end{quote}
\bigskip

\section{Introduction\label{sec:introduction}}

A useful tool for studying hadronic  decay amplitudes of charmed mesons is approximate 
flavor SU(3) symmetry. First order symmetry breaking corrections in amplitudes, due to 
the light quark mass term in the Standard Model Lagrangian, are expected to 
be of order $(m_s - m_{u,d})/\Lambda_{\rm QCD} \sim f_K/f_\pi -1\sim 0.2-0.3$. 

In the seventies, shortly after the discovery of charm, SU(3) group theory has been 
used to obtain amplitude relations for charmed meson decays
into pairs involving two pseudoscalar mesons or a pseudoscalar and a vector 
meson~\cite{Kingsley:1975fe}. This study was extended in the early nineties to 
include numerous first order SU(3) breaking terms one of which was assumed 
to dominate over the others~\cite{Savage:1991wu,Pirtskhalava:2011va}. A diagrammtic 
approach~\cite{Chau:1986jb}, equivalent to SU(3) group theory, has been developed 
and applied to fit data for branching ratios as they have been accumulated~\cite{Chiang:2002mr}. Assumptions made in these studies about SU(3) breaking were often model-dependent. Other  
studies of these decay processes went much beyond SU(3) by assuming factorization of hadronic 
amplitudes~\cite{Buccella:1990sp}. A recent SU(3) fit to current data of 
charmed meson decays to two pseudoscalar mesons worked reasonably well when 
including two of the numerous first order SU(3) breaking terms (of order $30\%$) available in 
a group theoretical approach~\cite{Hiller:2012xm}. 

While flavor SU(3), including $30\%$ first order symmetry breaking, has been shown to 
describe adequately a vast amount of data of hadronic decays of charmed mesons, 
this has not provided a precision test. For this matter one would hope to develop a 
perturbative expansion in reasonably 
small SU(3) breaking parameters, in which only high order symmetry breaking terms survive certain relations among amplitudes. A small step in this direction was made in 
Ref.\,\cite{Grossman:2012ry}, searching for linear relations among amplitudes for two-body and quasi two-body charmed meson decays in which first order SU(3) breaking terms cancel. Testing these relations, expected to hold within several percent, requires in most cases measuring relative strong phases between amplitudes which is a highly challenging task.

The purpose of this paper is to review, expand and improve new results obtained in recent work 
published in two short reports~\cite{Gronau:2013xba,Gronau:2014nda} applying a high order perturbation expansion in flavor SU(3) breaking and treating carefully isospin breaking. 
We will identify a class of $D^0$ decays, for which high-precision nonlinear relations among magnitudes of amplitudes hold. The lowest order symmetry breaking terms affecting these 
relations will be shown to be fourth order SU(3) breaking terms, and terms which are first order 
in isospin breaking and second order in SU(3) breaking \cite{isospinsecond}. 

One major motivation for this work is searching for signals of new physics. 
Very precise relations as discussed here, which would fail at some high order flavor 
symmetry breaking, could provide such signatures. For great convenience we will use 
U-spin, an SU(2) subgroup of flavor SU(3), rather than applying the full SU(3) group structure. 
Our U-spin expansion is, of course, consistent with SU(3) 
expansions in Refs.~\cite{Pirtskhalava:2011va,Hiller:2012xm,Grossman:2012ry}.

In Section \ref{sec:U=1pairs} we identify sets of four two-particle final states in $D^0$ 
decays, each consisting  of pairs involving charged pseudoscalar and vector mesons. 
In a given set two of these states and a linear combination of the other two form a 
U-spin triplet, playing an important role in $D^0$--$\bar D^0$ mixing. Our discussion 
in Sections \ref{sec:Usymmetry} - \ref{sec:isospin} studies in detail one of these sets 
denoted by $D^0\to P^+P^-$ consisting of $D^0 \to \pi^+K^-, K^+\pi^-, 
K^+K^-, \pi^+\pi^-$.  We derive U-spin symmetry relations for these processes in Section 
\ref{sec:Usymmetry}, and study first order and arbitrary order U-spin breaking corrections
in Sections \ref{sec:first} and \ref{sec:n}, respectively. In Section \ref{sec:relation} we 
obtain a high-precision relation, obeyed up to fourth order U-spin breaking, for ratios of 
amplitudes of the above four processes. Section 
\ref{sec:isospin} investigates first order isospin breaking terms occurring in this relation, 
showing that they are also suppressed by factors associated with second order U-spin breaking.
Sections \ref{sec:expPP}, \ref{sec:expV+P-} and \ref{sec:expP+V-} discuss experimental
tests in $D^0 \to P^+P^-$, $D^0 \to V^+P^-$ and $D^0\to P^+V^-$, respectively, where 
$V^\pm$ denote charged vector mesons. In Section \ref{sec:neutrals} we study $D^0$ 
decays into pairs of neutral pseudoscalar mesons, deriving amplitude relations and
$K^0_S-K^0_L$ rate asymmetry relations which hold up to second order U-spin breaking. 
A short conclusion is given in Section \ref{sec:conclusion}.

\section{$D^0$ decays to two-body $U=1$ states\label{sec:U=1pairs}}

An SU(2) subgroup of flavor SU(3) that is useful for studying charmed mesons is 
U-spin~\cite{Meshkov:1964zz}. The quark pair $(d, s)$ behaves like a doublet 
under this group while the $u$ quark and the heavier $c, b$ and $t$ quarks are 
singlets. U-spin symmetry leads to interesting relations among amplitudes of 
hadronic $D$ decays~\cite{Kingsley,Gronau:2000ru}. It also implies the vanishing of 
$D^0$--$\bar D^0$ mixing up to second order U-spin breaking~\cite{Gronau:2012kq},
which underlies the vanishing of $D^0$--$\bar D^0$ mixing within full 
flavor SU(3)~\cite{Falk:2001hx}. This behavior of $D^0$--$\bar D^0$ mixing under U-spin has 
been shown to follow from a cancellation up to second order U-spin breaking of 
mixing contributions from intermediate U-spin triplet states.  
The high order U-spin breaking perturbation expansion that will be studied in this 
paper works well, as we will show, for these two-body or quasi two-body $D^0$ decays
to $U=1$ states.

One class of $U=1$ two-body states involves pairs of opposite charge 
pseudoscalar or vector mesons. Our study will focus on these final states for which
a high-order U-spin breaking expansion is applicable. Another class of processes 
involves decay into pairs of neutral mesons. In this case two-body final states do not 
have well-defined values of U-spin. Instead, linear superpositions of final states have 
$U=1$. Consequently in these decays a U-spin expansion works well for certain linear 
combinations of decay amplitudes.

We start by classifying single meson states of positive or negative charge as  
doublets of U-spin. Since a pair of $d$ and $s$ quark and their antiquarks form two 
U-spin doublets, $(d, s)$ and $(\bar s, -\bar d)$, one has two doublets of pseudoscalar 
mesons 
\beq
P^+ = \left( \begin{array}{c} K^+ \cr 
-\pi^+\end{array}\right) \equiv  \left( \begin{array}{c} u\bar s \cr 
-u \bar d \end{array}\right)~,~~P^- = 
\left( \begin{array}{c} \pi^- \cr K^- \end{array} \right) \equiv
\left( \begin{array}{c} d \bar u \cr s \bar u \end{array} \right)~,
\eeq
and two doublets of vector mesons
\beq
 V^+ = \left( \begin{array}{c} K^{*+} \cr -\rho^+ \end{array}
\right)~,~~V^- = \left( \begin{array}{c} \rho^- \cr K^{*-} \end{array} \right)~.
\eeq
One can then form two-particle states of charge zero in four different forms, $P^+P^-, 
V^+P^-$, $P^+V^-$ and $V^+V^-$. 

In the next several sections we will study the four processes 
$D^0 \to P^+P^-,~P^+P^-=\pi^+K^-, K^+\pi^-$, $\pi^+\pi^-$, $K^+K^-$, for which 
most precise data exist. This discussion is also applicable to the other three sets of 
processes, $D^0 \to V^+P^-$, $P^+V^-$ and $D^0\to V^+V^-$.
For $D^0 \to V^+V^-$ one may, in principle, treat separately S, P and D-wave amplitudes,
or amplitudes for longitudinal polarizations, and for tranverse polarizations which are
mutually parallel and perpendicular to each other.  We will not discuss further these latter
challenging decay modes for which no branching ratios have been 
measured \cite{Agashe:2014kda}. 
  
 \section{U-spin symmetry limit\label{sec:Usymmetry}}

As a starting point we derive amplitude relations in the U-spin symmetry 
limit. The four possible two-particle states $|P^+P^-\rangle$ can 
be written in the form of three U-spin triplet states and one singlet state:
\beq\label{U=1}
|K^+\pi^-\rangle = |1, 1\rangle\,,~~~~
|\pi^+K^-\rangle = -|1,-1\rangle\,,~~~~
\frac{1}{\s} |K^+K^- - \pi^+\pi^- \rangle = |1,0\rangle\,,
\eeq
\beq\label{U=0}
\frac{1}{\s} |K^+K^- + \pi^+\pi^- \rangle = |0,0\rangle~.
\eeq

The charm-changing weak Hamiltonian has a simple transformation property under U-spin.
Its three pieces responsible for Cabibbo-favored (CF), singly Cabibbo-suppressed (SCS) and 
doubly Cabibbo-suppressed (DCS) decays transform,  when normalized suitably, like three 
components of a U-spin triplet operator denoted $(U=1, U_3=-1, 0, +1)$: 
\bea\label{H}
H^{\rm CF}_W & = & \frac{G_F}{\s}\cos^2\theta_C(\bar s\,c)(\bar u\,d) = -\cos^2\theta_C(1,-1)~,
\nonumber\\
H^{\rm SCS}_W & = & \frac{G_F}{\s}\cos\theta_C\sin\theta_C[(\bar s\,c)(\bar u\,s) - (\bar d\,c)(\bar u\,d)]
= \s \cos\theta_C\sin\theta_C(1,0)~,
\nonumber\\
H^{\rm DCS}_W & = & -\frac{G_F}{\s}\sin^2\theta_C(\bar d\,c)(\bar u\,s) = -\sin^2\theta_C(1,+1)~.
\eea
We have suppressed the chiral structure of V-A operators, using
$V_{ud}=V_{cs}=\cos\theta_C, V_{us}=-V_{cd}=\sin\theta_C$ for Cabibbo-Kobayashi-Maskawa 
(CKM) matrix elements.  

Charmed meson decays are dominated by real CKM matrix
elements associated with the first two quark families. For the most part we will be using 
a parametrization of the CKM matrix up to terms which are fourth order in $\lambda \equiv
V_{us}$  \cite{Agashe:2014kda}. In Section \ref{sec:relation} we will discuss the effect 
of higher order terms in $\lambda$ on four ratios of amplitudes studied in this section. 

Virtual $b$ quarks in penguin amplitudes may lead to tiny direct CP asymmetries in SCS 
decays of order $(\alpha_s(m^2_c)/\pi)(|V_{cb}V_{ub}|/|V_{cs}V_{us}|)
\sim 10^{-4}$~\cite{Brod:2011re}, depending on the final state. No CP asymmetries 
of this order are expected in CF and DCS decays. An order of magnitude larger 
asymmetries may occur in SCS decays which are subject to potential 
penguin amplitude enhancement~\cite{Golden:1989qx}. No CP asymmetries at these 
small levels have been measured so far. We will seek amplitude relations which hold 
at this high precision but not at higher accuracy. Errors in amplitudes are half the 
errors measured in decay rates. Thus we will neglect in our discussion direct CP asymmetries,
assuming that branching ratios for $D^0\to f$ are given by averages measured for decay 
processes and their CP conjugates,
\beq
{\cal B}(D^0 \to f) = {\cal B}(D^0 \to f)_{\rm CPav} \equiv \half[{\cal B}(D^0 \to f) + 
{\cal B}(\bar D^0 \to \bar f)]~.
\eeq 

The $D^0$ is a U-spin singlet. Denoting hadronic matrix elements in the U-spin symmetry limit 
by a superscript $(0)$ and defining a reduced matrix element,
$A \equiv \langle 1, U_3|$ $(1, U_3)|0,0\rangle$, one obtains from Es.\,(\ref{U=1}) and (\ref{H})  
\bea\label{pi+K-}
 \frac{\langle \pi^+K^-|H^{\rm CF}_W|D^0\rangle^{(0)}}{\cos^2\theta_C} & = & A~,
 \\
 \label{KKpipi}
\frac{\langle K^+K^- - \pi^+\pi^-|H^{\rm SCS}_W|D^0\rangle^{(0)}}{\cos\theta_C\sin\theta_C}
& = & 2A~,
\\
\label{K+pi-}
\frac{\langle K^+\pi^-|H^{\rm DCS}_W|D^0\rangle^{(0)}}{-\sin^2\theta_C} & = & A~.
\eea
Eq.\,(\ref{U=0}) leads to
\beq 
\langle K^+K^- + \pi^+\pi^-|H_W^{\rm SCS}|D^0\rangle^{(0)} \propto 
\langle 0, 0|(1, 0)|0, 0\rangle = 0~.
\eeq
Using (\ref{KKpipi}) this implies
\beq\label{KK}
\frac{\langle K^+K^-|H^{\rm SCS}_W|D^0 \rangle^{(0)}}{\cos\theta_C\sin\theta_C} = 
\frac{\langle \pi^+\pi^-|H^{\rm SCS}_W|D^0 \rangle^{(0)}}{-\cos\theta_C\sin\theta_C} =  A~.
\eeq
Thus the four amplitudes in (\ref{pi+K-}), (\ref{K+pi-}) and (\ref{KK}) denoted by the decay 
final state, $A(f)\equiv \langle f|H_W|D^0\rangle$
have simple ratios in the U-spin symmetry limit~\cite{Kingsley:1975fe},
\beq
A^{0)}(\pi^+K^-) : A^{(0)}(K^+K^-) : A^{(0)}(\pi^+\pi^-) : A^{(0)}(K^+\pi^-) = 
1 : \tan\theta_C : -\tan\theta_C : - \tan^2\theta_C~.
\eeq

We note that a derivation of the two ratios, $A^{0)}(K^+\pi^-)/A^{(0)}(\pi^+K^-) 
=-\tan^2\theta_C$ and $A^{0)}(\pi^+\pi^-)/A^{(0)}(K^+K^-)$  $=-1$, uses only  symmetry 
under $d\leftrightarrow s$ reflection~\cite{Gronau:2000ru} implying, 
\bea
\langle \pi^+K^-|(\bar s c)(\bar u d)|D^0\rangle & = &
 \langle K^+\pi^-|(\bar d c)(\bar u s)|D^0\rangle~,
 \nonumber\\
 \langle K^+ K^-|(\bar s c)(\bar u s) - (\bar d c)(\bar u d)|D^0\rangle & = &
 - \langle \pi^+ \pi^-|(\bar s c)(\bar u s) - (\bar d c)(\bar u d)|D^0\rangle~,
\eea
 and does not require full SU(2) U-spin symmetry. This full symmetry is required for the 
 relation between these two pairs of processes. 
 
 \section{First order U-spin breaking\label{sec:first}}
 
First order U-spin breaking corrections to the amplitudes (\ref{pi+K-}), (\ref{K+pi-}) and 
(\ref{KK}) are obtained by multiplying the weak Hamiltonian or the final states by an $s$$-$$d$ spurion mass operator,
$M_{\rm Ubrk} \propto (\bar s s) - (\bar d d)$~\cite{Savage:1991wu}:
\beq\label{first}
\langle f|H_{\rm eff}|D^0\rangle^{(1)} = \langle f|H_{\rm eff}M_{\rm Ubrk}|D^0\rangle + 
\langle M_{\rm Ubrk}f|H_{\rm eff}|D^0\rangle~.
\eeq
While for CF and DCS decays one has simply 
\beq\label{CFDCSfirst}
H_{\rm eff}^{\rm CF, DCS}M_{\rm Ubrk} = H_W^{\rm CF, DCS}M_{\rm Ubrk}~,
\eeq
the effective Hamiltonian for SCS decays obtains at first order an additional 
nonperturbative $s+d$ penguin term $P_{s+d}$ due to an $s-d$ mass 
difference~\cite{Bhattacharya:2012ah}: 
\beq\label{SCSfirst}
H_{\rm eff}^{\rm SCS}M_{\rm Ubrk} =  H_W^{\rm SCS}M_{\rm Ubrk}  + P_{s+d}~.
\eeq
The $U=0$ penguin amplitude in SCS decays interferes with opposite signs with the 
$U=1$ tree amplitudes in $D^0\to K^+K^-$ and in $D^0\to \pi^+\pi^-$. This  may potentially
increase the first amplitude and decrease the second. This effect of the penguin 
amplitude has been pointed out very early in 
Refs.\,\cite{Suzuki:1979uf,Abbott:1979fw,Savage}, and 
has been studied recently in  Refs.\,\cite{Bhattacharya:2012ah,Feldmann:2012js,Brod:2012ud} 
with its implication on CP asymmetries in these processes.

Eqs.\,(\ref{CFDCSfirst}) and (\ref{SCSfirst}) imply different first order U-spin breaking
corrections in $D^0 \to \pi^+K^-$ and $D^0 \to K^+\pi^-$, on the one hand, and in 
$D^0\to K^+K^-$ and $D^0\to \pi^+\pi^-$, on the other. We will show now that the
corrections within each of these two pairs of processes have equal magnitudes and 
opposite signs when normalized by the corresponding U-spin symmetric amplitudes.
We will first follow a full SU(2) U-spin argument presented in \cite{Gronau:2013xba}, and
then derive this result in a simpler manner using  $d\leftrightarrow s$ reflection.

Consider first Eq.\,(\ref{first}) for decays to the two U-spin triplet states, $|f_1 \rangle=
|\pi^+K^- \rangle = -|1,-1\rangle$ and $|f_2 \rangle = |K^+\pi^-\rangle = |1, 1\rangle$,
to which (\ref{CFDCSfirst}) applies. $H_W^{\rm CF}$ and $H_W^{\rm DCS}$ transform like
$(1, -1)$ and $(1, +1)$ while the $s$$-$$d$ spurion mass operator $M_{\rm Ubrk}$ behaves like $(1, 0)$. Since the $D^0$ is a U-spin singlet only the triplet operators  in the products  
$H^{\rm DCS, CF}_{\rm eff}M_{\rm Ubrk} \propto (1,\pm1)(1,0)$ 
contribute to the triplet final states $\pm|1,\pm 1\rangle$, and only the triplet 
states in $M_{\rm Ubrk}|1,\pm 1\rangle \propto (1,0)|1, \pm 1\rangle$ obtain  
contributions from the triplet Hamiltonian operator. Thus the two terms in (\ref{first}) involve 
coupling of the product $(1, \pm 1)\otimes (1, 0)$ into $(1, \pm 1)$, where $\pm$ signs 
correspond to $K^+\pi^-$ and $\pi^+ K^-$ final states.  Using a property of Clebsch-Gordan 
coefficients,
 \beq\label{CG}
(1,1;n,0|1,1) = (-1)^n(1,-1;n,0|1,-1)~,
\eeq
we therefore find
\beq\label{firstnorm}
\frac{\langle\pi^+ K^-|H_{\rm eff}^{\rm CF}|D^0\rangle^{(1)}}{\cos^2\theta_C} 
= \frac{\langle K^+ \pi^-|H_{\rm eff}^{\rm DCS}|D^0\rangle^{(1)}}{\sin^2\theta_C}~.
\eeq
That is, first order U-spin breaking terms contribute equally but 
with opposite signs to $D^0\to \pi^+K^-$ and $D^0\to K^+\pi^-$,
when normalized by corresponding U-spin symmetric amplitudes
given in Eqs.\,(\ref{pi+K-}) and (\ref{K+pi-}):
\beq\label{eps1}
\frac{\langle\pi^+ K^-|H_{\rm eff}^{\rm CF}|D^0\rangle^{(1)}}
{\langle\pi^+ K^-|H_W^{\rm CF}|D^0\rangle^{(0)}}
= -\frac{\langle K^+ \pi^-|H_{\rm eff} ^{\rm DCS}|D^0\rangle^{(1)}}
{\langle K^+ \pi^-|H_W^{\rm DCS}|D^0\rangle^{(0)}}\equiv - \epsilon_1~.
\eeq
We denote by $\epsilon_1$ the U-spin breaking term in $D^0\to K^+\pi^-$ 
normalized by its U-spin invariant amplitude. 
 
A simpler and rather immediate derivation of (\ref{eps1}) may be obtained by 
applying $d \leftrightarrow s$ reflection to Eq.\,(\ref{first}) for $f=\pi^+K^-$. Noting 
that the $s-d$ spurion mass operator $M_{\rm Ubrk}$ changes sign under this 
reflection, one has
\bea\label{minus}
& & \langle \pi^+K^-|(\bar s c)(\bar u d)M_{\rm Ubrk}|D^0\rangle + 
\langle M_{\rm Ubrk}(\pi^+K^-)|(\bar s c)(\bar u d)|D^0\rangle 
\nonumber\\
= & & =
-\langle K^+\pi^-|(\bar d c)(\bar u s)M_{\rm Ubrk}|D^0\rangle -
\langle M_{\rm Ubrk}(K^+\pi^-)|(\bar d c)(\bar u s)|D^0\rangle~.
\eea
This leads directly to (\ref{firstnorm}) and (\ref{eps1}).
 
This short argument applies also to first order U-spin breaking in SCS decays, 
$D^0\to K^+K^-$ and $D^0\to \pi^+\pi^-$, since the penguin operator $P_{s+d}$
in (\ref{SCSfirst}) also changes sign under $d\leftrightarrow s$ reflection.
Therefore,
\beq\label{eps2}
\frac{\langle\pi^+ \pi^-|H_{\rm eff}^{\rm SCS}|D^0\rangle^{(1)}}
{\langle\pi^+ \pi^-|H_W^{\rm SCS}|D^0\rangle^{(0)}}
= -\frac{\langle K^+ K^-|H_{\rm eff} ^{\rm SCS}|D^0\rangle^{(1)}}
{\langle K^+ K^-|H_W^{\rm SCS}|D^0\rangle^{(0)}}\equiv - \epsilon_2~,
\eeq
where $\epsilon_2$ denotes the U-spin breaking term in $D^0\to K^+K^-$ 
normalized by its U-spin invariant amplitude.
Furthermore, the change in sign between first order terms in $D^0\to K^+K^-$ 
and $D \to \pi^+\pi^-$ applies separately to contributions of the two 
operators on the right-hand side of (\ref{SCSfirst}) representing tree and 
penguin amplitudes. This leads one to expect that the U-spin breaking parameter 
$\epsilon_2$, involving both tree and penguin amplitudes, is larger than $\epsilon_1$
which involves only tree amplitudes~\cite{Bhattacharya:2012ah}.

Combining the results (\ref{eps1}) and (\ref{eps2}) with the zeroth order Eqs.\,(\ref{pi+K-}), 
(\ref{K+pi-}) and (\ref{KK}) one obtains the following first order expressions for decay 
amplitudes:
\bea\label{ampfirst}
A(D^0\to \pi^+ K^-) & = & \cos^2\theta_C\,A(1-\epsilon_1)~,
\nonumber\\
A(D^0 \to K^+\pi^-) & = & -\sin^2\theta_C\,A(1 + \epsilon_1)~,
\nonumber\\
A(D^0 \to \pi^+\pi^-) & = & -\cos\theta_C\sin\theta_C\,A(1 - \epsilon_2)~,
\nonumber\\
A(D^0 \to K^+K^-) & = & \cos\theta_C\sin\theta_C\,A(1 + \epsilon_2)~.
\eea

\section {U-spin breaking of arbitrary order\label{sec:n}}
 
 U-spin breaking of order $n$ in decay amplitudes $\langle f|H_{\rm eff}|D^0\rangle$ is 
 obtained by introducing in the Hamiltonian or in the final state a total of $n$ powers of the 
 $s-d$ spurion mass operator, applying (\ref{SCSfirst}) to SCS decays. Generalizing the 
 argument for a relative negative sign in first order breaking, based on a change of sign of 
 $M_{\rm Ubrk}$ and $P_{s+d}$ under $d \leftrightarrow s$ reflection, we conclude that
 a negative relative sign applies to odd $n$ and a positive sign to even $n$:
 \beq\label{nKpinorm}
\frac{\langle\pi^+ K^-|H_{\rm eff}^{\rm CF}|D^0\rangle^{(n)}}{\cos^2\theta_C} 
= (-1)^n\frac{\langle K^+ \pi^-|H_{\rm eff}^{\rm DCS}|D^0\rangle^{(n)}}{-\sin^2\theta_C}~.
\eeq
\beq\label{nKKnorm}
\frac{\langle K^+K^-|H^{\rm SCS}_W|D^0 \rangle^{(n)}}{\cos\theta_C\sin\theta_C} = (-1)^n
\frac{\langle \pi^+\pi^-|H^{\rm SCS}_W|D^0 \rangle^{(n)}}{-\cos\theta_C\sin\theta_C}~.
\eeq
Thus we may sum up: 
 \bea\label{4amps}
A(D^0\to \pi^+ K^-) & = & \cos^2\theta_C\,A[1-\epsilon_1 + a_1\epsilon_1^2 - 
a'_1\epsilon_1^3 +...]~,
\nonumber\\
A(D^0 \to K^+\pi^-) & = & -\sin^2\theta_C\,A[1 + \epsilon_1 + a_1\epsilon_1^2 + 
a'_1\epsilon_1^3 +...]~,
\nonumber\\
A(D^0 \to \pi^+\pi^-) & = & -\cos\theta_C\sin\theta_C\,A[1 - \epsilon_2 + a_2\epsilon_2^2 
-a'_2\epsilon_2^3 + ...]~,
\nonumber\\
A(D^0 \to K^+K^-) & = & \cos\theta_C\sin\theta_C\,A[1 + \epsilon_2 + a_2\epsilon_2^2 
+ a'_2\epsilon_2^3 + ...]~.
\eea
While the complex U-spin breaking parameters $\epsilon_{1,2}$ and the nonerturbative
coefficients $a_{1,2}, a'_{1,2},...$ are not calculable from first principles, one expects
the first two parameters to be around $0.2-0.3$ ($\epsilon_2$ being larger in magnitude 
than $\epsilon_1$ for the above-mentioned argument) and the coefficients to be of order 
one, $|a_{1,2}|\sim |a'_{1,2}|\sim 1$.

Expanding magnitudes of amplitudes up to and including third order we find 
\bea\label{expansion}
|1 \pm \epsilon + a\epsilon^2 \pm a'\epsilon^3| = 1 & \pm & {\rm Re}\,\epsilon + 
\hf({\rm Im}\,\epsilon)^2 + {\rm Re}(a\epsilon^2) \pm {\rm Re}(a'\epsilon^3)
\nonumber\\
& \mp & \hf{\rm Re}\,\epsilon\,({\rm Im}\,\epsilon)^2 
\pm {\rm Im}\,\epsilon\,{\rm Im}(a\epsilon^2)~.
\eea
In the next section we will use this expansion for studying ratios of magnitudes of decay amplitudes, identifying relations among ratios in which U-spin breaking terms up to third order 
cancel, thus being sensitive to tiny fourth order U-spin breaking.  We will then argue that  
there is no need to go beyond third order in (\ref{expansion}) for showing that fourth order 
terms do not cancel in these relations. 

\section{High-precision relation among ratios of amplitudes\label{sec:relation}}

 We define four ratios of amplitudes:
 \bea\label{Ri}
 R_1  & \equiv & \frac{|A(D^0 \to K^+\pi^-)|}{|A(D^0 \to \pi^+K^-)|\tan^2\theta_C}~,
 \nonumber\\
 R_2 & \equiv & \frac{|A(D^0 \to K^+K^-)|}{|A(D^0\to \pi^+\pi^-)|}~,
 \nonumber\\
 R_3 & \equiv & \frac{|A(D^0 \to K^+K^-)| + |A(D^0\to \pi^+\pi^-)|}
{|A(D^0 \to \pi^+K^-)|\tan\theta_C + |A(D^0\to K^+\pi^-)|\tan^{-1}\theta_C}~, 
\nonumber\\
R_4 & \equiv & \sqrt{\frac{|A(D^0 \to K^+K^-)||A(D^0\to \pi^+\pi^-)|}
{|A(D^0\to \pi^+K^-)||A(D^0\to K^+\pi^-)|}}~.
\eea
These four ratios are not mutually independent. 
They obey a trivial identity,
\beq\label{identity}
R_4 = R_3 \sqrt{\frac{1 - [(R_2-1)/(R_2+1)]^2}{1 - [(R_1-1)/(R_1+1)^2}}~.
\eeq
We will prove a nontrivial relation involving $R_3$ and $R_4$ and another nonlinear 
function of $R_1$ and $R_2$, that holds up to {\em maximal} fourth order U-spin breaking.  

Using (\ref{expansion}) we start by expanding the two ratios $R_1$ and $R_2$ up to 
third order U-spin breaking:
\bea\label{R12}
R_1 & = & 1 + 2[{\rm Re}\,\epsilon_1 + ({\rm Re}\,\epsilon_1)^2] + 
{\cal O}(\epsilon_1^3)~,
\nonumber\\
R_2 & = & 1 + 2[{\rm Re}\,\epsilon_2 + ({\rm Re}\,\epsilon_2)^2] +
 {\cal O}(\epsilon_2^3)~.
\eea
These two ratios involve first order corrections given by $2{\rm Re}\,\epsilon_1$ 
and $2{\rm Re}\,\epsilon_2$. Second order corrections in these ratios are 
given by squares of these same real parts with no dependence on the unknown 
parameters  $a_1$ and $a_2$. Thus measurements of $R_1$ and $R_2$ 
provide a way for calculating ${\rm Re}\,\epsilon_1$ and ${\rm Re}\,\epsilon_2$
up to third order corrections.
Solutions for ${\rm Re}\,\epsilon_1$ and ${\rm Re}\,\epsilon_2$ using Eqs.\,(\ref{R12}) 
should include the U-spin symmetry limit, requiring ${\rm Re}\,\epsilon_1=0$ 
and ${\rm Re}\,\epsilon_2=0$ for $R_1=1$ and $R_2=1$, respectively, rather 
than ${\rm Re}\,\epsilon_1=-1$ and ${\rm Re}\,\epsilon_2=-1$.  This implies
\bea\label{Re}
{\rm Re}\,\epsilon_1 & = & \hf\left(\sqrt{2R_1 - 1} -1\right) +  {\cal O}(\epsilon_1^3)~,
\nonumber\\
{\rm Re}\,\epsilon_2  & = & \hf\left(\sqrt{2R_2 - 1} -1\right) + {\cal O}(\epsilon_2^3)~.
\eea
As we will see immediately, these two first order U-spin breaking parameters do not 
determine only $R_1$ and $R_2$ but  also the difference $R_3-R_4$.

The two ratios $R_3$ and $R_4$, in which first 
and third order terms cancel  \cite{GrosRob,MGJR}, may be expanded up to fourth order:
\bea\label{R34}
R_3 & = & 1 + \hf[({\rm Im}\,\epsilon_2)^2 - ({\rm Im}\,\epsilon_1)^2] + 
{\rm Re}\,[a_2\epsilon_2^2 - a_1\epsilon_1^2] + {\cal O}(\epsilon_1^4, \epsilon_2^4)~,
\nonumber\\
R_4 & = & 1 - \hf{\rm Re}\,(\epsilon_2^2 - \epsilon_1^2) + 
{\rm Re}\,(a_2\epsilon_2^2 - a_1\epsilon_1^2) + {\cal O}(\epsilon_1^4, \epsilon_2^4)
\nonumber \\
& = & 1 + \hf[({\rm Im}\,\epsilon_2)^2 - ({\rm Im}\,\epsilon_1)^2] + 
{\rm Re}\,(a_2\epsilon_2^2 - a_1\epsilon_1^2) 
- \hf[({\rm Re}\,\epsilon_2)^2 - ({\rm Re}\,\epsilon_1)^2]
\nonumber \\
& & + ~{\cal O}(\epsilon_1^4, \epsilon_2^4)~.
\eea
These two ratio are noticed to differ by second order U-spin breaking terms depending
solely on ${\rm Re}\,\epsilon_1$ and ${\rm Re}\,\epsilon_2$:
\beq\label{R3-4}
R_3 - R_4 = \hf[({\rm Re}\,\epsilon_2)^2 - ({\rm Re}\,\epsilon_1)^2]
+ {\cal O}(\epsilon_1^4, \epsilon_2^4)~.
\eeq
Using
\beq
({\rm Re}\,\epsilon_i)^2 = \frac{1}{4}\left(\sqrt{2R_i - 1} -1\right)^2 + 
2{\rm Re}\,\epsilon_i\,{\cal O}(\epsilon_i^3) = 
\frac{1}{4}\left(\sqrt{2R_i - 1} -1\right)^2 +  {\cal O}(\epsilon_i^4)~,~~~i=1,2~,
\eeq
one obtains the the following nonlinear relation among the four ratios of amplitudes, which 
holds up to tiny fourth order U-spin breaking terms:
\beq\label{NLrel}
\Delta R \equiv R_3 - R_4 + \frac{1}{8}\left[\left(\sqrt{2R_1-1} - 1\right)^2 - 
\left(\sqrt{2R_2 - 1} -1\right )^2\right]=
{\cal O}(\epsilon_1^4, \epsilon_2^4)~.
\eeq

This relation may also be obtained by expanding the identity (\ref{identity}) to first order in
$[(R_{1,2}-1)/(R_{1,2} +1)]^2$ and applying (\ref{R12}) and (\ref{Re}).
Fourth order terms in $({\rm Re}\,\epsilon_i)^2$ are proportional to ${\rm Re}\,\epsilon_i$.
It can be  easily shown that this is not the case for terms of this order occurring in $R_3$, 
$R_4$ and in their difference. Therefore one concludes that fourth order terms do not 
cancel in $\Delta R$ and in any higher order expansion of the square root in (\ref{identity}). 
Thus, in hindsight, there is no way of obtaining a U-spin breaking relation of higher order 
than (\ref{NLrel}) and one does not need to go beyond third order in (\ref{expansion}). 

In our derivation of (\ref{NLrel}) we have used a parametrization of the CKM matrix 
up to terms which are fourth order in $\lambda \equiv V_{us}= 0.2253 \pm 0.008$ 
\cite{Agashe:2014kda}. We now study the effects  of higher order terms in 
$\lambda$ on $R_i$ and on this relation. Including $\lambda^4$ and $\lambda^5$ 
terms in the CKM matrix one has~\cite{Antonelli:2009ws}
\bea
V_{ud} & = & 1 - \hf \lambda^2 - \frac{1}{8}\lambda^4~,
\nonumber\\
V_{us} & = & \lambda~,
\nonumber\\
V_{cd} & = & -\lambda + \hf A^2\lambda^5[1 - 2(\rho + i\eta)]~,
\nonumber\\
V_{cs } & = &  1 - \hf \lambda^2 - \frac{1}{8}\lambda^4 - \hf A^2\lambda^4~,
\eea
where $A = 0.82 \pm 0.02, \rho = 0.12 \pm 0.02, 
\eta = 0.36 \pm 0.02$ \cite{Agashe:2014kda}. 

We define a small parameter $\xi$ determining the effect of $\lambda^4$ and $\lambda^5$ 
terms on ratios of amplitudes:
\beq
1 + \xi \equiv \frac{1 - \hf A^2\lambda^4|1 - 2(\rho + i\eta)|}
{1 - \hf A^2\lambda^4}~.
\eeq
Using the above values of $\lambda, A, \rho$ and $\eta$ we find $|\xi| \lesssim 10^{-4}$. 
While $R_4$ is unaffected by $\xi$, $R_3$ and the two quadratic terms
in (\ref{NLrel}) obtain corrections suppressed by $\xi$ and by first order U-spin breaking.
These corrections, respectively $-\hf\xi({\rm Re}\,\epsilon_1+ {\rm Re}\,\epsilon_2)$,  
$+\hf\xi{\rm Re}\,\epsilon_1$ and $+\hf\xi{\rm Re}\,\epsilon_2$, cancel each other. The 
remaining corrections in (\ref{NLrel}), suppressed by $\xi$ and by second order U-spin 
breaking, are much below the level of $10^{-4}$ and may be safely neglected relative 
to tiny fourth order U-spin breaking terms of order $10^{-3}$ or $10^{-4}$.

Taking $\epsilon_i\sim 0.2$ as a typical value for first order U-spin breaking, the nonlinear 
relation $\Delta R=0$ is expected to hold at a very high precision of order 
$\epsilon_i^4\sim 10^{-3}$. We will confirm this prediction in Section \ref{sec:expPP}. 
At this high precision one cannot ignore first order isospin breaking terms, which one 
would generally assume to be around $(m_d - m_u)\Lambda_{\rm QCD} \sim 10^{-2}$.
We will study these terms in the next section showing that, in fact, they are suppressed 
by both isospin breaking parameters and by second order U-spin breaking.

 \section{First order isospin breaking\label{sec:isospin}}
 
First order isospin breaking is introduced by multiplying the weak Hamiltonian by 
a $d-u$ spurion mass operator,
\beq\label{isobreak}
M_{\rm Ibrk} \propto (\bar d d - \bar u u) =  \hf(\bar d d + \bar s s) 
- \bar u u + \hf(\bar d d - \bar s s)~,
\eeq 
transforming like a combination of a U-spin singlet and triplet. Isospin breaking 
contributions of the U-spin singlet operator in the four amplitudes (\ref{4amps}) 
are identical when normalized by suitable CKM factors, and may be absorbed 
into the U-spin symmetric amplitude $A$. 
This is true also for $U=0$ isospin breaking electromagnetic interactions because 
the $d$ and $s$ quarks have identical charges.

Contributions of the triplet operator in (\ref{isobreak})
follow the signs of first order U-spin breaking corrections. They are represented by two 
distinct first order isospin breaking parameters, $\delta_1$ - for U-spin triplet states 
$\pi^+K^-$ and $K^+\pi^-$, and $\delta_2$ - for $K^+K^-$ and $\pi^+\pi^-$, the two 
component of a U-spin singlet state. Thus
\bea\label{4ampsdel}
A(D^0\to \pi^+ K^-) & = & \cos^2\theta_C\,A(1-\epsilon_1 + a_1\epsilon_1^2 - 
a'_1\epsilon_1^3 -\delta_1 +...)~,
\nonumber\\
A(D^0 \to K^+\pi^-) & = & -\sin^2\theta_C\,A(1 + \epsilon_1 + a_1\epsilon_1^2 + 
a'_1\epsilon_1^3 +\delta_1 +...)~,
\nonumber\\
A(D^0 \to \pi^+\pi^-) & = & -\cos\theta_C\sin\theta_C\,A(1 - \epsilon_2 + a_2\epsilon_2^2 +
-a'_2\epsilon_2^3 -\delta_2 + ...)~,
\nonumber\\
A(D^0 \to K^+K^-) & = & \cos\theta_C\sin\theta_C\,A(1 + \epsilon_2 + a_2\epsilon_2^2 +
+ a'_2\epsilon_2^3 +\delta_2 + ...)~.
\eea
 
Instead of (\ref{expansion}) we now expand: 
\bea\label{expansion_isospin}
|1 \pm \epsilon + a\epsilon^2 \pm a'\epsilon^3 \pm \delta| & = & 1 \pm  {\rm Re}\,\epsilon + 
\hf({\rm Im}\,\epsilon)^2 + {\rm Re}(a\epsilon^2) \pm {\rm Re}(a'\epsilon^3)
\nonumber\\
& \mp & \hf{\rm Re}\,\epsilon\,({\rm Im}\,\epsilon)^2 
\pm {\rm Im}\,\epsilon\,{\rm Im}(a\epsilon^2) \pm {\rm Re}\,\delta + 
{\rm Im}\,\delta\,{\rm Im}\,\epsilon~.
\eea
This characteristic amplitude expansion includes two new terms, $\pm {\rm Re}\,\delta$ and 
$+{\rm Im}\,\delta\,{\rm Im}\,\epsilon$, the latter involving suppression by both isospin 
and U-spin breaking parameters. We will show that terms of this order do not affect the nonlinear relation (\ref{NLrel}).
 
The expansion of the four ratios of amplitudes now 
includes terms which are first order in isospin breaking and other terms suppressed by 
both isospin and U-spin breaking:
\bea\label{R12del}
R_1 & = & 1 + 2[{\rm Re}\,\epsilon_1 + ({\rm Re}\,\epsilon_1)^2] + 2{\rm Re}\,\delta_1 
+ 4{\rm Re}\,\delta_1\,{\rm Re}\,\epsilon_1 + {\cal O}(\epsilon_1^3) + 
{\cal O}(\delta_1\epsilon_1^2)~,
\nonumber\\
R_2 & = & 1 + 2[{\rm Re}\,\epsilon_2 + ({\rm Re}\,\epsilon_2)^2] + 2{\rm Re}\,\delta_2 
+ 4{\rm Re}\,\delta_2\,{\rm Re}\,\epsilon_2 + {\cal O}(\epsilon_2^3) +
 {\cal O}(\delta_2\epsilon_2^2)~,
\eea
\bea\label{R34del}
R_3 & = & 1 + \hf[({\rm Im}\,\epsilon_2)^2 - ({\rm Im}\,\epsilon_1)^2] + 
{\rm Re}\,(a_2\epsilon_2^2 - a_1\epsilon_1^2) 
+({\rm Im}\,\delta_2\,{\rm Im}\,\epsilon_2 - {\rm Im}\,\delta_1\,{\rm Im}\,\epsilon_1)
\nonumber\\
& & +~{\cal O}(\epsilon_{1,2}^4) 
+ {\cal O}(\delta_{1,2}\epsilon_{1,2}^3)~,
\nonumber\\
R_4 & = & 1 + \hf[({\rm Im}\,\epsilon_2)^2 - ({\rm Im}\,\epsilon_1)^2] + 
{\rm Re}\,(a_2\epsilon_2^2 - a_1\epsilon_1^2) 
+({\rm Im}\,\delta_2\,{\rm Im}\,\epsilon_2 - {\rm Im}\,\delta_1\,{\rm Im}\,\epsilon_1)
\nonumber\\
& & -~\hf[({\rm Re}\,\epsilon_2)^2 - ({\rm Re}\,\epsilon_1)^2]
- ({\rm Re}\,\delta_2{\rm Re}\,\epsilon_2 - {\rm Re}\,\delta_1{\rm Re}\,\epsilon_1) 
+ {\cal O}(\epsilon_{1,2}^4) + {\cal O}(\delta_{1,2}\epsilon_{1,2}^3)\,.
\eea
Eqs.\,(\ref{R12del}) imply for $i=1,2$
\beq\label{Reidel}
{\rm Re}\, \epsilon_i = \hf \left(\sqrt{2R_i-1}-1\right) - {\rm Re}\delta_i - 
2{\rm Re}\,\delta_i{\rm Re}\,\epsilon_i + {\cal O}(\delta_i\epsilon_i) + 
{\cal O}(\epsilon_i^3)~,
\eeq
or
\beq
\left({\rm Re}\,\epsilon_i \right)^2 = \frac{1}{4}\left(\sqrt{2R_i - 1} - 1\right)^2 
-2{\rm Re}\,\delta_i{\rm Re}\,\epsilon_i + {\cal O}(\delta_i\epsilon_i^2) 
+{\cal O}(\epsilon_i^4)~.
\eeq
Eqs.\,(\ref{R34del}) lead to
\beq
R_3 - R_4 = \hf \left[({\rm Re}\,\epsilon_2)^2 - {\rm Re}\,\epsilon_1)^2\right]
+({\rm Re}\,\delta_2{\rm Re}\,\epsilon_2  - {\rm Re}\,\delta_1{\rm Re}\,\epsilon_1) + 
{\cal O}(\epsilon_{1,2}^4) + {\cal O}(\delta_{1,2}\epsilon_{1,2}^3)~.
\eeq 
Consequently isospin breaking terms of the form ${\rm Re}\,\delta_2{\rm Re}\,\epsilon_2 -
{\rm Re}\,\delta_1{\rm Re}\,\epsilon_1$ cancel in $\Delta R$:
\beq\label{NLreldel}
\Delta R \equiv R_3 - R_4 + \frac{1}{8}\left[\left(\sqrt{2R_1-1} - 1\right)^2 - 
\left(\sqrt{2R_2 - 1} -1\right )^2\right] =
{\cal O}(\epsilon_1^4, \epsilon_2^4) + {\cal O}(\delta_1\epsilon_1^2, \delta_2\epsilon_2^2)~.
\eeq

It is remarkable that isospin breaking modifies the nonlinear relation (\ref{NLrel}) by 
terms which are suppressed by both first order isospin breaking and second order U-spin 
breaking factors. Taking  $\delta_i\sim 10^{-2}$, $\epsilon_i\sim 0.2-0.3$, these terms are expected 
to be at most $10^{-3}$, similar in magnitude to fourth order U-spin breaking terms affecting this 
relation.  

\section{Experimental tests in $D^0\to P^+P^-$\label{sec:expPP}}

We will now apply current experimental data to study the hierarchy among U-spin 
breaking terms of increasing order. Our final goal is testing the predicted amplitude 
relation~(\ref{NLreldel}). 
Hadronic decay amplitudes ($A$) are obtained from measured branching ratios 
(${\cal B}$) by eliminating phase space factors depending on final particles 
center-of-mass 3-momenta ($p$), and on the $D$ meson mass and its lifetime 
($M_D$ and $\tau_D$), 
\beq
|A| = M_D\sqrt{\frac{8\pi{\cal B}}{\tau_D\,p}}~.
\eeq
In our calculation of amplitudes we will disregard a common factor $M_D\sqrt{8\pi/\tau_D}$ 
which cancels in the four ratios $R_i$. Values for measured branching 
ratios~\cite{Agashe:2014kda}, center-of-mass momenta, and amplitudes defined in this 
manner are quoted in Tables \ref{tab:PP}. 
 
Note that all four amplitudes include a factor ${\cal B}_{\pi K}^{1/2}$ corresponding to 
the branching fraction of the Cabibbo-favored decay $D^0\to \pi^+K^-$.
We have included no error in the amplitude for this process 
as the other three branching ratios (including errors) have been measured relative 
to this process~\cite{Agashe:2014kda}. The relative errors in the amplitudes of these three
processes are all below the level of one percent. We will assume no correlation between 
these errors, which have been measured in three independent analyses for different 
final states. The high precision achieved recently by the CDF, LHCb and Belle 
collaborations in measuring the DCS amplitude is 
remarkable~\cite{Aaltonen:2013pja}, as it required time-dependent separation between 
this highly suppressed decay and $D^0$--$\bar D^0$ mixing followed by the CF decay.

\begin{table}[t]
\caption{Branching fractions and amplitudes for $D^0\to P^+P^-$ 
decays~\cite{Agashe:2014kda}\label{tab:PP}} 
\begin{center}
\begin{tabular}{c c c c } \hline \hline
Decay mode  & Branching fraction & $p$ (${\rm GeV}/c$) & 
$|A|=\sqrt{{\cal B}/p}\,({\rm GeV}/c)^{-1/2}$  \\ \hline
 $D^0\to \pi^+K^-$  &  ${\cal B}_{\pi\hskip-0.5mmK}=
 (3.88\pm 0.05)\hskip-1mm\times\hskip-1mm10^{-2}$  & $0.861$ & 
 $1.078{\cal B}_{\pi\hskip-0.5mmK}^{1/2}$ \\
 $D^0 \to K^+\pi^- $ &  $(3.56 \pm 0.06)\hskip-1mm\times\hskip-1mm10^{-3}
 {\cal B}_{\pi\hskip-0.5mmK}$ & $0.861$ & $(0.06430\pm 0.00054)
 {\cal B}_{\pi\hskip-0.5mmK}^{1/2}$ \\ 
 $D^0 \to \pi^+\pi^-$ & $(3.59\pm 0.06)\hskip-1mm\times\hskip-1mm10^{-2}
 {\cal B}_{\pi\hskip-0.5mmK}$ & $0.922$ & 
 $(0.1973\pm 0.0016){\cal B}_{\pi\hskip-0.5mmK}^{1/2}$ \\ 
 $D^0 \to K^+K^-$ & $(10.10 \pm 0.16)\hskip-1mm\times\hskip-1mm10^{-2}
 {\cal B}_{\pi\hskip-0.5mmK}$ & $0.791$ & 
 $(0.3573\pm 0.0028){\cal B}_{\pi\hskip-0.5mmK}^{1/2}$ \\
 \hline\hline
 \end{tabular}
\end{center}
\end{table}

Using values of amplitudes given in Tables \ref{tab:PP} and 
$\tan\theta_C=0.23125 \pm 0.00082$ \cite{Agashe:2014kda} we calculate the four 
ratios $R_i$ defined in Eq.\,(\ref{Ri}),
\bea\label{Rinum}
R_1 & = & 1.115 \pm 0.012~,
\nonumber\\
R_2  & = & 1.811 \pm 0.020~,
\nonumber\\
R_3 & = & 1.052 \pm 0.008~,
\nonumber\\
R_4 & = & 1.008 \pm 0.007~.
\eea
Errors in the ratios have been obtained by adding in quadrature errors in the relevant 
amplitudes.  

We have seen that the first two ratios involve first order U-spin breaking terms. These terms,
depending on two distinct U-spin breaking parameters $\epsilon_1$ and $\epsilon_2$,  
are considerably larger in $R_2$ than in $R_1$. This has been anticipated in the discussion  
below Eq.\,(\ref{eps2}). Using (\ref{Reidel}), in which we neglect first order isospin breaking and third order U-spin breaking, we calculate reasonably small U-spin breaking 
parameters,
\bea\label{Renum}
{\rm Re}\,\epsilon_1 & = &0.054 \pm 0.005~,
\nonumber\\
{\rm Re}\,\epsilon_2 & = & 0.310  \pm 0.006~.
\eea
The other two ratios, $R_3$ and $R_4$, given in (\ref{R34del}) in terms of $\epsilon_{1,2}$ 
and coefficients $a_{1,2}$ of order one, deviate from one by second order U-spin breaking 
terms (we neglect terms suppressed by both isospin and U-spin breaking and fourth order 
terms in U-spin breaking),
\bea\label{secondorder}
R_3 - 1 =  \hf[({\rm Im}\,\epsilon_2)^2 - ({\rm Im}\,\epsilon_1)^2] + 
{\rm Re}\,(a_2\epsilon_2^2 - a_1\epsilon_1^2) & = & 0.052 \pm 0.08~,
\nonumber\\
R_4 - 1 = {\rm Re}\,[(a_2 - \hf)\epsilon_2^2 - (a_1- \hf)\epsilon_1^2]  & = & 0.008 \pm 0.007~.
\eea 
{\em The hierarchy between the first order parameters in (\ref{Renum}) and the second order 
terms calculated in (\ref{secondorder}) confirms and justifies the perturbative approach 
we have applied in this study to U-spin breaking.} Without having a method for calculating 
the nonperturbative coefficients $a_i$, the almost exact cancellation of second order terms 
in $R_4$ seems to be accidental. In view of the small value of ${\rm Re}\,\epsilon_1$ 
and the much larger value of ${\rm Re}\,\epsilon_2$ this approximate cancellation seems 
to imply $a_2\simeq 1/2$. 

Having shown that second order U-spin breaking terms are a few percent, one 
expects fourth order terms to be of order $10^{-3}$, comparable in magnitude or larger 
than terms which are first order in isospin breaking and second order in U-spin breaking. 
Let us now check this prediction in the relation (\ref{NLreldel}) which contains on the right-hand 
side terms of these two kinds. Using the definitions of $R_i$ in (\ref{Ri}) and adding in quadrature 
errors in amplitudes [rather than errors in $R_i$ given in (\ref{Rinum})], we obtain
\beq
\Delta R = (-3.2 \pm 0.4)\times 10^{-3}~.
\eeq
This confirms our prediction.

\section{Experimental tests in $D^0\to V^+P^-$\label{sec:expV+P-}}

As mentioned, the discussion in Sections \ref{sec:Usymmetry} - \ref{sec:isospin} applies 
also to three other classes of processes involving one or two charged vector mesons, 
$D^0\to V^+P^-, D^0\to P^+V^-$ and $D^0\to V^+V^-$. In particular, a nonlinear amplitude 
relation similar to (\ref{NLreldel}) holds in each one of these classes with a precision 
depending on the size of U-spin breaking. In this section we summarize concisely the 
situation relevant to this question in $D^0 \to V^+P^-$, consisting of the four processes, 
$D^0 \to \rho^+ K^-, D^0 \to K^{*+}\pi^-, D^0 \to \rho^+\pi^-$ and $D^0\to K^{*+}K^-$. 
We denote first order U-spin breaking and isospin breaking parameters in these processes
by $\epsilon'_{1,2}$ and $\delta'_{1,2}$, respectively,  in analogy to $\epsilon_{1,2}$ and 
$\delta_{1,2}$ in $D^0 \to P^+P^-$. 

Defining four ratios of amplitudes $R'_i$ in analogy with (\ref{Ri}),
 \bea\label{Ri'}
 R'_1  & \equiv & \frac{|A(D^0 \to K^{*+}\pi^-)|}{|A(D^0 \to \rho^+K^-)|\tan^2\theta_C}~,
 \nonumber\\
 R'_2 & \equiv & \frac{|A(D^0 \to K^{*+}K^-)|}{|A(D^0\to \rho^+\pi^-)|}~,
 \nonumber\\
 R'_3 & \equiv & \frac{|A(D^0 \to K^{*+}K^-)| + |A(D^0\to \rho^+\pi^-)|}
{|A(D^0 \to \rho^+K^-)|\tan\theta_C + |A(D^0\to K^{*+}\pi^-)|\tan^{-1}\theta_C}~, 
\nonumber\\
R'_4 & \equiv & \sqrt{\frac{|A(D^0 \to K^{*+}K^-)||A(D^0\to \rho^+\pi^-)|}
{|A(D^0\to \rho^+K^-)||A(D^0\to K^{*+}\pi^-)|}}~,
\eea
one obtains a sum rule analogous to (\ref{NLreldel}):
\beq\label{NLreldel'}
\Delta R' \equiv R'_3 - R'_4 + \frac{1}{8}\left[\left(\sqrt{2R'_1-1} - 1\right)^2 - 
\left(\sqrt{2R'_2 - 1} -1\right )^2\right] =
{\cal O}(\epsilon'^4_1, \epsilon'^4_2) + 
{\cal O}(\delta'_1\epsilon'^2_, \delta'_2\epsilon'^2_2)~.
\eeq

Magnitudes of amplitudes for the P-wave decays $D^0 \to V^+P^-$ are obtained from corresponding branching ratios using
\beq
|A| = M_D\sqrt{\frac{8\pi{\cal B}}{\tau_D\,p^3}}~.
\eeq
We will disregard again the factor $M_D\sqrt{8\pi/\tau_D}$ since we are only concerned 
with ratios of amplitudes.
Values for measured branching ratios~\cite{Agashe:2014kda}, center-of-mass momenta, 
and amplitudes defined in this manner are quoted in Tables \ref{tab:VP}. Relative 
errors in CF and SCS amplitudes are reasonably small, between two and three percent. 
In contrast, the relative error in the DCS amplitude $|A(D^0\to K^{*+}\pi^-)|$, obtained 
by the CLEO and BaBar collaborations
through Dalitz plot analyses of $D^0\to K_S\pi^+\pi^-$~\cite{Asner:2003uz}, is quite 
large,  $^{+26\%}_{-15\%}$. This large asymmetric error limits considerably the precision 
of $R'_1, R'_3$ and  $R'_4$. Adding in quadrature errors in relevant amplitudes, measured independently for different three-body final states, we calculate
\bea
R'_1 & = & 0.971^{+0.257}_{-0.148}~,
\nonumber\\
R'_2 & = & 0.939 \pm 0.029~,
\nonumber\\
R'_3 & = & 1.061^{+0.082}_{-0.140}~,
\nonumber\\
R'_4 & = & 1.061^{+0.083}_{-0.142}~.
\eea

%
\begin{table}[t]
\caption{Branching fractions and amplitudes for $D^0\to V^+P^-$ 
decays~\cite{Agashe:2014kda}\label{tab:VP}} 
\begin{center}
\begin{tabular}{c c c c } \hline \hline
Decay mode  & Branching fraction & $p$ (${\rm GeV}/c$) & 
$|A|=\sqrt{{\cal B}/p^3}\,({\rm GeV}/c)^{-3/2}$ \\ \hline
 $D^0 \to \rho^+K^-$ & $0.108 \pm 0.007$ & $0.675$ & $0.593 \pm 0.019$ \\
 $D^0 \to K^{*+}\pi^-$ & $(3.42^{+1.80}_{-1.02})\hskip-1mm\times\hskip-1mm 
 10^{-4}$ & $0.711$ & $0.0308^{+0.0081}_{-0.0046}$  \\
 $D^0 \to \rho^+\pi^-$ & $(9.8 \pm 0.4)\hskip-1mm\times\hskip-1mm 10^{-3}$ 
 & $0.764$ & $0.148 \pm 0.003$ \\
 $D^0 \to K^{*+}K^-$ & $(4.38 \pm 0.21)\hskip-1mm\times\hskip-1mm 10^{-3}$ 
 & $0.610$ & $0.1389 \pm  0.0033$ \\
 \hline \hline
\end{tabular}
\end{center}
\end{table}

Expansions similar to (\ref{R12del}) and (\ref{R34del}) apply to these four ratios in 
terms of $\epsilon'_{1,2}$ and $\delta'_{1,2}$. The leading U-spin breaking corrections 
in $R'_1$ and $R'_2$ are first order, while in $R'_3$ and $R'_4$ they are second order. 
The measured values of the first two ratios imply
\bea\label{Re'}
{\rm Re}\,\epsilon'_1& = & \hf\left(\sqrt{2R'_1 - 1} -1\right)  = -0.015^{+0.118}_{-0.083}~,
\nonumber\\
{\rm Re}\,\epsilon'_2 & = & \hf\left(\sqrt{2R'_2 - 1} -1\right)  =  -0.032 \pm 0.016~.
\eea 
We note that {\em the numerical value of ${\rm Re}\,\epsilon'_2$ is smaller by about an 
order of magnitude than the value of ${\rm Re}\,\epsilon_2$} calculated in (\ref{Renum}). 
That is, first order U-spin breaking in SCS $D^0 \to V^+P^-$ decays is about 
an order of magnitude smaller than in corresponding $D^0\to P^+P^-$ decays. 
This seems to imply that an enhancement of the U-spin breaking penguin amplitude 
suggested to occur in $D^0\to P^+P^-$~\cite{Golden:1989qx} is not at work in 
$D^0\to V^+ P^-$.  
The other U-spin breaking parameter, ${\rm Re}\,\epsilon'_1$, does not 
involve a penguin amplitude. In spite of the current large error in this parameter 
the first of Eqs.\,(\ref{Re'}) favors strongly $|{\rm Re}\,\epsilon'_1|\le 0.1$, suggesting 
that a value close to that measured for $|{\rm Re}\,\epsilon_1|$ is not unlikely. 

These values of ${\rm Re}\,\epsilon'_{1,2}$ imply that typical second order U-spin 
breaking terms in $R'_{3,4}$ are around one percent. Confirming this prediction
and obtaining a more precise value for 
${\rm Re}\,\epsilon'_1$ requires a substantial improvement in the measurement of 
${\cal B}(D^0 \to K^{*+}\pi^-)$. In the meantime we use the values measured for 
$|{\rm Re}\,\epsilon'_{1,2}|$ to argue that fourth order U-spin breaking terms should be 
around $10^{-4}$. This is also expected to be the magnitude of terms in $\Delta R'$ 
suppressed by both isospin breaking and by second order U-spin breaking. 
Thus, due to smaller U-spin breaking parameters in $D^0\to V^+P^-$ relative to $D^0\to P^+P^-$ 
one predits $\Delta R'$ to be about an order of magnitude smaller than $\Delta R$.
Using the amplitudes in Table \ref{tab:VP} we calculate
\beq\label{DeltaR'}
\Delta R' = (0.2 ^{+3.2}_{-5.5})\times 10^{-4},
\eeq
where the error is dominated by the error in $|A(D^0 \to K^{*+}\pi^-)|$. 
This confirms our prediction. A more precise test could be achieved by improving 
the measurement of the branching ratio for this process.  
 
\section{Experimental tests in $D^0 \to P^+V^-$\label{sec:expP+V-}}
%
\begin{table}[t]
\caption{Branching fractions and amplitudes for $D^0\to P^+V^-$ 
decays~\cite{Agashe:2014kda}\label{tab:PV}} 
\begin{center}
\begin{tabular}{c c c c } \hline \hline
Decay mode  & Branching fraction & $p$ (${\rm GeV}/c$) & 
$|A|=\sqrt{{\cal B}/p^3}\,({\rm GeV}/c)^{-3/2}$ \\ \hline
$D^0 \to \pi^+ K^{*-}$ & $(4.98^{+0.45}_{-0.51})\hskip-1mm\times\hskip-1mm 10^{-2}$ 
& $0.711$ & $0.372^{+0.017}_{-0.019}$ \\
$D^0 \to K^+ \rho^-$ & -- & $0.675$ & -- \\
 $D^0 \to \pi^+\rho^-$ & $(4.96 \pm 0.24)\hskip-1mm\times\hskip-1mm 10^{-3}$ 
 & $0.764$ & $0.1055 \pm 0.0026$  \\
 $D^0 \to K^+ K^{*-}$ & $(1.56 \pm 0.12)\hskip-1mm\times\hskip-1mm 10^{-3}$ 
 & $0.610$ & $0.0829 \pm  0.0032$ \\
 \hline \hline
\end{tabular}
\end{center}
\end{table}

Current branching ratios and amplitudes for $D^0 \to P^+V^-$, consisting of 
$D^0 \to \pi^+ K^{*-}, K^+ \rho^-$, $\pi^+ \rho^-, K^+ K^{*-}$ are given in Table III.  
Ref.\,\cite{Agashe:2014kda}\label{tab:PV} quotes no branching ratio measurement for 
the DCS decay $D^0 \to K^+\rho^-$. Three-body decays, $D^0 \to K^+ \pi^- \pi^0$, 
involving $K^+\rho^-, K^{*+}\pi^-$ and other intermediate states, have been observed 
by Belle~\cite{Tian:2005ik} and BaBar \cite{Aubert:2006kt} with branching ratios 
$(3.18 \pm 0.29)\times 10^{-4}$ and $(2.97 \pm 0.19)\times 10^{-4}$, respectively. 
The $D^0 \to K^+\pi^-\pi^0$ events involve interference of DCS decays with 
$D^0$--$\bar D^0$ mixing followed by CF decays.  Evidence for $D^0$--$\bar D^0$ mixing 
at 3.2 standard deviation was presented by Babar  \cite{Aubert:2008zh}, measuring the 
fraction of $K^+\rho^-$ in these events to be $(39.8\pm 6.5)\%$. (No interference would 
have implied ${\cal B}(D^0 \to K^+\rho^-) \sim 1.2 \times 10^{-4}$.) More work is needed 
for resolving the effect of $D^0$--$\bar D^0$ mixing on these events, and for obtaining 
a solid measurement of ${\cal B}(D^0 \to K^+\rho^-)$.

We define ratios of amplitudes $R''_i$ in analogy with (\ref{Ri}),
 \bea\label{Ri'}
 R''_1  & \equiv & \frac{|A(D^0 \to K^+\rho^-)|}{|A(D^0 \to \pi^+K^{*-})|\tan^2\theta_C}~,
 \nonumber\\
 R''_2 & \equiv & \frac{|A(D^0 \to K^+K^{*-})|}{|A(D^0\to \pi^+\rho^-)|}~,
 \nonumber\\
 R''_3 & \equiv & \frac{|A(D^0 \to K^+K^{*-})| + |A(D^0\to \pi^+\rho^-)|}
{|A(D^0 \to \pi^+K^{*-})|\tan\theta_C + |A(D^0\to K^+\rho^-)|\tan^{-1}\theta_C}~, 
\nonumber\\
R''_4 & \equiv & \sqrt{\frac{|A(D^0 \to K^+K^{*-})||A(D^0\to \pi^+\rho^-)|}
{|A(D^0\to \pi^+K^{*-})||A(D^0\to K^+\rho^-)|}}~.
\eea
We denote first order U-spin breaking and isospin breaking parameters in these 
amplitudes by $\epsilon''_{1,2}$ and $\delta''_{1,2}$, respectively. 

In the absence of a solid measurement of $|A(D^0\to K^+\rho^-)|$ one can only 
calculate $R''_2$. Neglecting isospin breaking and third order U-spin breaking one 
obtains
\beq
R''_2 = 1 + 2\left[{\rm Re}\,\epsilon''_2 + ({\rm Re}\,\epsilon''_2)^2\right] = 0.786 \pm 0.036~,
\eeq
which implies
\beq
{\rm Re}\,\epsilon''_2 = -0.122 \pm 0.024~.
\eeq
That is, the magnitude of the U-spin breaking parameter in SCS $D^0 \to P^+V^-$ decays is intermediary between corresponding parameters in $D^0 \to P^+P^-$ 
(${\rm Re}\,\epsilon_2 = 0.310 \pm 0.006$) and $D^0 \to V^+P^-$ 
(${\rm Re}\,\epsilon'_2= -0.032 \pm 0.016$). Namely, no significant 
U-spin breaking penguin enhancement applies to $D^0 \to P^+V^-$. 
One expects a similar or smaller magnitude for ${\rm Re}\,\epsilon''_1$. 

The two ratios $R''_3$ and $R''_4$ deviate from one by second order U-spin breaking 
terms [see Eqs.\, (\ref{R34})] which are expected to be at most a few percent. 
Using $R''_3 = 1 \pm 0.05$, where we include a conservative uncertainty 
of $5\%$ due to second order U-spin breaking corrections, one obtains 
the following prediction for  ${\cal B}(D^0\to K^+\rho^-)$ \cite{GRob}, 
\beq
|A(D^0 \to K^+\rho^-)|= 0.0237 \pm 0.0025~~~~\Rightarrow~~~
{\cal B}(D^0\to K^+\rho^-) = (1.73 \pm 0.36)\times 10^{-4}~.
\eeq
For comparison, assuming $R''_4 = 1 \pm 0.05$ implies a very similar prediction,
\beq\label{DCS}
|A(D^0 \to K^+\rho^-)| = 0.0235 \pm 0.0028~~~\Rightarrow~~~
{\cal B}(D^0\to K^+\rho^-) = (1.70  \pm 0.40)\times 10^{-4}~.
\eeq
This value of $|A(D^0 \to K^+\rho^-)|$ would imply ${\rm Re}\,\epsilon''_1 = 0.08\pm 0.06$,
comparable in magnitude to ${\rm Re}\,\epsilon''_2$ and in agreement with expectation.

Taking for $|A(D^0 \to K^+\rho^-)|$ the value in (\ref{DCS}), using the three 
measured amplitudes quoted in Table III, and assuming no error correlation 
between the four amplitudes, one obtains 
\beq\label{NLreldel''}
\Delta R'' \equiv R''_3 - R''_4 + \frac{1}{8}\left[\left(\sqrt{2R''_1-1} - 1\right)^2 - 
\left(\sqrt{2R''_2 - 1} -1\right )^2\right] = (0.8^{+2.2}_{-5.6})\times 10^{-4}~.
\eeq
This value, which is similar to (\ref{DeltaR'}), is in agreement with our prediction 
that, in view of the above values of ${\rm Re}\,\epsilon''_{1,2}$,
fourth order U-spin breaking terms and isospin breaking terms suppressed by 
second order U-spin breaking should be of order $10^{-4}$. 
The positive error in $\Delta R''$ is dominated by the uncertainty assumed in the 
unmeasured DCS amplitude. 
One still awaits a solid measurement of ${\cal B}(D^0\to K^+\rho^-)$ 
which would test the prediction (\ref{DCS}).  
The larger negative error in $\Delta R''$, originating in errors on
the three measured amplitudes in Table III, may be reduced by improving the 
corresponding branching ratio measurements. 
   
\section{$D^0$ decays to pairs of neutral pseudoscalars\label{sec:neutrals}}

In the U-spin symmetry limit one obtains simple amplitude relations for $D^0$ decays
to pairs of light neutral pseudoscalar mesons \cite{neutrals}. We will go through the 
symmetry argument first, extending it to include U-spin breaking at arbitrary order
in CF and DCS decays. We will then demonstrate a few amplitude relations 
which hold up to second order U-spin breaking.
Other relations of this kind have been studied in Ref.\,\cite{Grossman:2012ry} in
the framework of flavor SU(3).

In the symmetry limit one neglects $\eta-\eta'$ mixing which is due to first order U-spin 
breaking represented by the spurion mass operator $M_{\rm Ubrk}$. 
Thus, we will write amplitudes for $\eta = \eta_8$ in our discussion of the symmetry limit, 
while $\eta_8$ will be used explicitly when introducing U-spin breaking.
When discussing amplitudes, rates and asymmetries for $\eta_8$ we will assume (as has 
been assumed in Ref.\,\cite{Grossman:2012ry}) knowledge of amplitudes including 
a relative phase for $\eta$ and $\eta'$ and favored values of the mixing angle. 
Taking 
$\eta = \eta_8\equiv (2s\bar s - u\bar u - d\bar d)/\sqrt6$, the following superpositions of 
single neutral particle states belong to a U-spin triplet,
\beq\label{1}
| K^0\rangle \equiv |d\bar s\rangle = |1, 1\rangle~,~~~
|\bar K^0\rangle  \equiv |s \bar d\rangle = -|1, -1\rangle~,~~~
\hf(\sqrt 3|\eta\rangle - |\pi^0\rangle) \equiv |s \bar s - d \bar d\rangle/\s = |1, 0\rangle~,
\eeq
while the orthogonal U-spin singlet is
\beq\label{0}
\hf(|\eta\rangle + \sqrt 3|\pi^0\rangle) \equiv |s \bar s + d \bar d -2u \bar u\rangle/\sqrt 6 \
 = |0, 0\rangle~.
\eeq
Two-particle states in an S-wave are obtained by symmetrizing products of single
particle states. 

Symmetrized products of two $U=1$ single-particle states consist of
$U=0$ and $U=2$ (with $U_3= \pm 1, 0$) states, while the product of two $U=0$ 
states is pure $U=0$. 
$D^0$ decay matrix elements of the $U=1$ weak 
Hamiltonian (\ref{H}) vanish for each one of these five states. This implies the 
following five U-spin symmetry relations: (For short notation we denote amplitudes 
by their final states.)
\bea\label{vanish}
\sqrt 3A^{(0)}(K^0\eta) - A^{(0)}(K^0\pi^0)  & = & 0~,
\nonumber\\
\sqrt 3A^{(0)}(\bar K^0\eta) - A^{(0)}(\bar K^0\pi^0) & = & 0~,
\nonumber\\
A^{(0)}(K^0 \bar K^0) & = & 0~,
\nonumber\\
A^{(0)}(\eta\eta) + A^{(0)}(\pi^0\pi^0) & = & 0~,
\nonumber\\
\sqrt 3A^{(0)}(\eta\pi^0) +\s A^{(0)}(\pi^0\pi^0) & = & 0~.
\eea 
Hadronic matrix elements for symmetrized products of $U=1, U_3=\pm 1, 0$ and 
$U=0$ states are then given by a single $U=1$ amplitude ${\cal A}$:
\bea\label{Uneutral1}
\sqrt 3A^{(0)}(K^0\eta) = A^{(0)}(K^0\pi^0) & = & -\sin^2\theta_C\,{\cal A}~,
\\
\label{Uneutral2}
\sqrt 3A^{(0)}(\bar K^0\eta) = A^{(0)}(\bar K^0\pi^0) & = & \cos^2\theta_C\,{\cal A}~,
\\
\label{Uneutral3}
\sqrt 3A^{(0)}(\eta\pi^0) = \s A^{(0)}(\eta\eta) = -\s A^{(0)}(\pi^0\pi^0) & = &
\s\cos\theta_C\sin\theta_C\,{\cal A}~.
\eea 
Note that since we symmetrized final states also for identical particles, corresponding 
amplitudes have been divided by $\s$ in order that their squares give decay rates.

The above amplitude expressions are analogous to the U-spin symmetry 
expressions (\ref{pi+K-}), (\ref{K+pi-}) and (\ref{KK}) for decays into pairs of charged 
pseodoscalar mesons. Some of these relations follow merely from $d \leftrightarrow s$ 
symmetry. Since the $U=1$ and $U=0$ superpositions of $\pi^0$ and $\eta$ states in 
(\ref{1}) and (\ref{0}) are respectively antisymmetric and symmetric with respect to 
$d \leftrightarrow s$ reflection one has
\bea
\langle \bar K^0(\sqrt 3\eta - \pi^0) |(\bar s c)(\bar u d)|D^0\rangle 
&  = & - \langle K^0(\sqrt 3\eta - \pi^0) |(\bar d c)(\bar u s)|D^0\rangle~, 
\nonumber\\
\langle \bar K^0(\eta + \sqrt 3 \pi^0)|(\bar s c)(\bar u d)|D^0\rangle 
 & = &  \langle K^0(\eta + \sqrt 3 \pi^0)|(\bar d c)(\bar u s)|D^0\rangle~,
\eea  
implying
\bea\label{zero}
\frac{\sqrt 3 A^{(0)}(\bar K^0\eta) -A^{(0)}(\bar K^0\pi^0)}{\cos^2\theta_C}
& = & \frac{\sqrt 3 A^{(0)}(K^0\eta) - A^{(0)}(K^0\pi^0)}{\sin^2\theta_C} = 0~,
\nonumber\\
\frac{A^{(0)}(\bar K^0\eta) + \sqrt 3 A^{(0)}(\bar K^0\pi^0)}{\cos^2\theta_C}
& = & -\frac{A^{(0)}(K^0\eta) + \sqrt 3 A^{(0)}(K^0\pi^0)}{\sin^2\theta_C} = \frac{4}{\sqrt 3}{\cal A}~.
\eea
The right-hand sides follow using (\ref{Uneutral1}) and (\ref{Uneutral2}) based on 
the full SU(2) structure of U-spin.  

U-spin breaking of order $n$ in CF and DCS amplitudes is introduced  by 
multiplying transition operators or final states by a total of $n$ powers of 
the $s-d$ spurion mass operator  $M_{\rm Ubrk}$ which changes sign under 
$d \leftrightarrow s$. Consequently one has
\bea\label{Kpi_n}
\frac{\sqrt 3 A^{(n)}(\bar K^0\eta_8) -A^{(n)}(\bar K^0\pi^0)}{\cos^2\theta_C}
& = & (-1)^n\frac{\sqrt 3 A^{(n)}(K^0\eta_8) - A^{(n)}(K^0\pi^0)}{\sin^2\theta_C}~,
\nonumber\\
\frac{A^{(n)}(\bar K^0\eta_8) + \sqrt 3 A^{(n)}(\bar K^0\pi^0)}{\cos^2\theta_C}
& = & (-1)^n\frac{A^{(n)}(K^0\eta_8) + \sqrt 3 A^{(n)}(K^0\pi^0)}{-\sin^2\theta_C}~.
\eea
Denoting first order U-spin breaking parameters in these two pairs of processes by 
$\epsilon_0$ and $\epsilon'_0$, one may expand the above four linear combinations of 
amplitudes to arbitrary order,
\bea\label{neutralexpansion1}
\sqrt 3 A(\bar K^0\eta_8) - A(\bar K^0\pi^0) & = &
\cos^2\theta_C{\cal A}[ \epsilon_0 - a_0\epsilon_0^2 + ...]~,
\nonumber\\
\sqrt 3 A(K^0 \eta_8) - A(K^0\pi^0) & = &
-\sin^2\theta_C{\cal A}[ \epsilon_0 + a_0\epsilon_0^2 + ...]~,
\\
\label{neutralexpansion2}
\frac{\sqrt 3}{4}[A(\bar K^0\eta_8) + \sqrt 3 A(\bar K^0\pi^0)] & = &
\cos^2\theta_C {\cal A}[1 - \epsilon'_0  + a'_0\epsilon'^2_0 +...]~,
\nonumber\\
\frac{\sqrt 3}{4}[A(K^0\eta_8) + \sqrt 3 A(K^0\pi^0)] & = & 
-\sin^2\theta_C {\cal A}[1 + \epsilon'_0 + a'_0\epsilon'^2_0 + ...]~,
\eea
where $|a_0| \sim , |a'_0| \sim 1$. 

Eqs.\,(\ref{neutralexpansion1}) imply a linear amplitude relation in which first 
order U-spin breaking terms cancel,
\beq\label{neutrel}
[\sqrt 3 A(\bar K^0\eta_8) - A(\bar K^0\pi^0]\tan^2\theta_C + 
\sqrt 3 A(K^0 \eta_8) - A(K^0\pi^0) = 0~.
\eeq
This relation has also been obtained using a general first order SU(3) breaking 
expansion\,\cite{Grossman:2012ry}, in which a dozen SU(3) breaking parameters 
contributing to these processes cancel in this relation. 
We note that, while it follows from Eqs.\,(\ref{neutralexpansion2}) that  the linear relation
\beq
[A(\bar K^0\eta_8) + \sqrt 3 A(\bar K^0\pi^0)]\tan^2\theta_C - 
[A(K^0\eta_8) + \sqrt 3 A(K^0\pi^0)] = \frac{4}{\sqrt 3}\sin^2\theta_C {\cal A}
\eeq
is also free of first order U-spin breaking, the U-spin invariant amplitude ${\cal A}$ 
on the rignt-hand side is not necessarily invariant under the full flavor SU(3) group
when including other states. This is just like the amplitude $A$ defined above 
(\ref{pi+K-}) for decays to charged particles.  

The four amplitudes for two body decays involving $K^0$ or $\bar K^0$ and $\pi^0$ 
or $\eta_8$ have the following first order expansions:
\bea\label{K0first}
A(\bar K^0\pi^0) & = & \cos^2\theta_C{\cal A}\left[1 - \epsilon'_0 - \frac{1}{4}\epsilon_0 +
{\cal O}(\epsilon^2_0, \epsilon'^2_0)\right ]~,
\nonumber\\
\sqrt 3 A(\bar K^0\eta_8) & = & \cos^2\theta_C{\cal A}\left[1 - \epsilon'_0 + \frac{3}{4}\epsilon_0 +
{\cal O}(\epsilon^2_0, \epsilon'^2_0)\right ]~,
\nonumber\\
A(K^0\pi^0) & = & -\sin^2\theta_C{\cal A}\left[1 + \epsilon'_0 - \frac{1}{4}\epsilon_0 +
{\cal O}(\epsilon^2_0, \epsilon'^2_0)\right ]~,
\nonumber\\
\sqrt 3 A(K^0\eta_8) & = & -\sin^2\theta_C{\cal A}\left[1 + \epsilon'_0 + \frac{3}{4}\epsilon_0 +
{\cal O}(\epsilon^2_0, \epsilon'^2_0)\right ]~.
\eea
This implies that the two ratios of amplitudes in (\ref{Uneutral1}) and (\ref{Uneutral2}), 
which are equal in the U-spin symmetry limit, are also equal when including first order 
U-spin breaking:
\beq\label{ReE0}
\frac{\sqrt 3|A(K^0\eta_8|}{|A(K^0\pi^0)|} = \frac{\sqrt 3|A(\bar K^0\eta_8)|}{|A(\bar K^0\pi^0)|} 
= 1 + {\rm Re}\,\epsilon_0 + {\cal O}(\epsilon_0^2, \epsilon'^2_0)~,
\eeq
\beq\label{eps'0}
\frac{|A(K^0\pi^0)|}{|A(\bar K^0\pi^0)|\tan^2\theta_C} = 
\frac{|A(K^0\eta_8)|}{|A(\bar K^0\eta_8)|\tan^2\theta_C} = 1 + 2{\rm Re}\,\epsilon'_0 
+ {\cal O}(\epsilon_0^2, \epsilon'^2_0)~. 
\eeq

Branching ratio measurements of  the above two DCS decay modes are not feasible 
because a final state neutral kaon is identified in a $K_S$ or a $K_L$ state. This 
involves an interference between CF and DCS decays. 
A method for measuring this interference has been proposed in 
Ref. \cite{Bigi:1994aw}. Defining a rate asymmetry between decays involving 
$K^0_S$ and $K^0_L$,
\beq
R(D^0, M^0) \equiv \frac{\Gamma(D^0 \to K^0_S M^0) - \Gamma(D^0 \to K^0_L M^0)}
{\Gamma(D^0 \to K^0_S M^0) + \Gamma(D^0 \to K^0_L M^0)}~,~~~~(M^0 = \pi, \eta, \eta')
\eeq
one obtains, to leading order in the ratio of DCS and CF amplitudes,
\beq\label{R}
R(D^0, M^0) = -\frac {2{\rm Re}\,[A(\bar K^0 M^0)A^*(K^0 M^0)]}{|A(\bar K^0 M^0)|^2}~.
\eeq

Eqs.\,(\ref{K0first}) predict equal asymmetries for $M^0=\pi$ and $M^0 = \eta_8$
up to second order U-spin breaking. The two asymmetries are given by
\beq\label{Reps}
R(D^0, \eta_8) = R(D^0, \pi^0) = 2\tan^2\theta_C\left [1 + 2 {\rm Re}\,\epsilon'_0 + 
{\cal O}(\epsilon_0^2, \epsilon'^2_0)\right ]~.
\eeq
Comparing Eqs.\,(\ref{eps'0}) and (\ref{Reps}) we find
\beq
\frac{|A(K^0 M^0)|}{|A(\bar K^0 M^0)|} = \hf R(D^0, M^0)\left[1 + 
{\cal O}(\epsilon_0^2, \epsilon'^2_0)\right]~,~~~~(M^0=\pi^0, \eta_8)~.
\eeq
That is, although the branching fractions for DCS decays $D^0 \to K^0\pi^0$ and 
$D^0 \to K^0\eta_8$ cannot be measured directly, they may be  obtained up to 
second order U-spin breaking corrections from corresponding CF branching 
fractions and $K^0_S-K^0_L$ asymmetries:
\beq\label{B(K0)}
{\cal B}(D^0 \to K^0 M^0) = \frac{1}{4}[R(D^0, M^0)]^2{\cal B}(D^0 \to \bar K^0 M^0)
\left[1 + {\cal O}(\epsilon_0^2, \epsilon'^2_0)\right]~,~~~~(M^0= \pi, \eta_8)~.
\eeq 

%
\begin{table}[t]
\caption{Branching fractions and amplitudes for $D^0$ decays to pairs of neutral 
pseudoscalar mesons~\cite{Agashe:2014kda}\label{tab:neutrals}} 
\begin{center}
\begin{tabular}{c c c c } \hline \hline
Decay mode  & Branching fraction & $p$ (${\rm GeV}/c$) & 
$|A|=\sqrt{{\cal B}/p}\,({\rm GeV}/c)^{-1/2}$  \\ \hline
 $D^0\to K^0_S\pi^0$  &  $(1.19 \pm 0.04)\times 10^{-2}$  
 & $0.860$ & $--$ \\
 $D^0\to K^0_L\pi^0$  & $(1.00 \pm 0.07)\times 10^{-2}$  & $0.860$ & 
 $--$ \\
 $D^0 \to \bar K^0\pi^0$ & $(2.29 \pm 0.07)\times 10^{-2~a}$ & $0.860$ & $0.163 \pm 0.002$ \\
 $D^0 \to \bar K^0\eta$ &  $(0.890 \pm 0.079)\times 10^{-2}$ & $0.772$ & 
 $0.107\pm 0.005$ \\ 
 $D^0 \to \pi^0\pi^0$ & $(8.20\pm 0.35)\times 10^{-4}$ & $0.923$ & 
 $0.0298\pm 0.0006$ \\ 
 $D^0 \to \eta \pi^0$ & $(6.8 \pm 0.7) \times 10^{-4}$ & 
 $0.846$ & $0.0284\pm 0.0014$ \\
 \hline\hline
 \end{tabular}
\end{center}
\leftline{$^a$ Branching ratio calculated as twice the average of ${\cal B}(D^0 \to K^0_S\pi^0)$
and ${\cal B}(D^0 \to K^0_L\pi^0)$.}
\end{table}

Table \ref{tab:neutrals} summarizes current relevant information on branching ratios 
and amplitudes for $D^0$ decays into pairs of neutral pseudoscalars. We do not include
the $\eta\eta$ mode (and decays involving the $\eta'$), as $D^0 \to \eta_8\eta_8$ would 
include $D^0 \to \eta'\eta'$ which has zero phase space.

An estimate of U-spin breaking is given by a ratio of SCS and CF decay amplitudes 
which equals one in the symmetry limit [see (\ref{Uneutral2}) (\ref{Uneutral3})],
\beq
\frac{|A(\pi^0\pi^0)|}{|A(\bar K^0\pi^0)|\tan\theta_C} - 1 = -0.21 \pm 0.02~,
\eeq
Another quantity measuring U-spin breaking is
\beq\label{Reep0}
\frac{\sqrt 3|A(\bar K^0\eta)|}{|A(\bar K^0\pi^0)|} - 1
= 0. 14 \pm 0.05~.
\eeq
In order to determine ${\rm Re}\,\epsilon_0$ from (\ref{ReE0}) one would 
have to know also $|A(\bar K^0\eta')|$ and the relative strong phase between this 
amplitude and $A(\bar K^0\eta)$, using a 
favored value for the $\eta$--$\eta'$ mixing angle~\cite{eta_mixing}. 
The other U-spin breaking parameter, ${\rm Re}\,\epsilon'_0$, 
is obtained from (\ref{Reps}) using a $K^0_S-K^0_L$ asymmetry
measurement by the CLEO collaboration~\cite{He:2007aj}, 
$R(D^0,\pi^0) = 0.108 \pm 0.035$,
\beq\label{Reep'0}
{\rm Re}\,\epsilon'_0 = \frac{R(D^0,\pi^0) - 2\tan^2\theta_C}{4\tan^2\theta_C} = 
0.00 \pm 0.16~.
\eeq 
Arguments favoring small U-spin breaking in the asymmetries $R(D^0, M^0)$
have been presented in Ref.\,\cite{Rosner:2006bw} adopting a diagrammatic flavor 
SU(3) approach. 

The numerical values of ${\rm Re}\,\ep'_0$ in (\ref{Reep'0}) and of the above other two 
measured U-spin breaking quantities imply that second order U-spin breaking terms in 
Eqs.\,(\ref{ReE0}) -- (\ref{B(K0)}) are at most of order several percent. Neglecting these 
contributions involves an approximation that is quantitatively similar to the one used to obtain 
(\ref{R}), where terms which are second order in the ratio of DCS and CF amplitudes have 
been neglected.

\section{Conclusion\label{sec:conclusion}}

We described a new approach to hadronic $D^0$ decay amplitudes applying a 
perturbative expansion in U-spin breaking parameters and 
treating isospin breaking carefully. We have identified a class of two-body and quasi 
two-body decays involving charged pseudoscalars ($P$) and vector mesons ($V$), 
for which in each case adequate hierarchies have been shown to occur between 
U-spin breaking terms of increasing order.  

Nonlinear amplitude relations were predicted in each one of three cases, 
$D^0\to P^+P^-$, $D^0 \to V^+P^-$ and $D^0 \to P^+V^-$, which hold up to fourth 
order U-spin breaking and isospin breaking terms suppressed also by second order 
U-spin breaking. The three predicted relations have been shown to hold experimentally 
at a very high precision varying between $10^{-3}$ and $10^{-4}$, in agreement with our 
estimates of high order terms. 
More precise tests require a first robust measurement of ${\cal B}(D^0 \to K^+\rho^-)$ 
and improving branching ratio measurements for $D^0$ decays to $K^{*+}\pi^-, \pi^+ K^{*-}, 
\pi^+\rho^-$ and $K^+ K^{*-}$. So far no unexpected flavor symmetry breaking down to this 
very low level has been found, which would indicate physics beyond the standard model. 
This provides useful constraints on new $|\Delta C|=1$ operators, potentially originating 
in new physics at energies much above a TeV \cite{Gronau:2014pra}. 

Finally, we also studied decays to two neutral 
pseudoscalar mesons, deriving much less precise amplitude relations and relations 
for rate asymmetries between decays involving $K^0_S$ and $K^0_L$, which hold 
up to second order U-spin breaking terms at a level of several percent.
\medskip

I wish to thank Yuval Grossman and Dean Robinson for useful communications.


\begin{thebibliography}{99}
%
\bibitem{Kingsley:1975fe} 
R.~L.~Kingsley, S.~B.~Treiman, F.~Wilczek and A.~Zee,
Phys.\ Rev.\ D {\bf 11}, 1919  (1975);
M.~B.~Voloshin, V.~I.~Zakharov and L.~B.~Okun,
JETP Lett.\  {\bf 21}, 183 (1975)
[Pisma Zh.\ Eksp.\ Teor.\ Fiz.\  {\bf 21}, 403 (1975)];
L.~L.~Wang and F.~Wilczek,
Phys.\ Rev.\ Lett.\  {\bf 43}, 816 (1979);
C.~Quigg,
Z.\ Phys.\ C {\bf 4}, 55 (1980).
%
%
\bibitem{Savage:1991wu}
  M.~J.~Savage,
  Phys.\ Lett.\ B {\bf 257}, 414 (1991);
%
  W.~Kwong and S.~P.~Rosen,
  Phys.\ Lett.\ B {\bf 298}, 413 (1993);
%
  I.~Hinchliffe and T.~A.~Kaeding,
  Phys.\ Rev.\ D {\bf 54}, 914  (1996)
  [hep-ph/9502275].
   %
\bibitem{Pirtskhalava:2011va}
For a recent study including CP asymmetries see
  D.~Pirtskhalava and P.~Uttayarat,
  Phys.\ Lett.\ B {\bf 712}, 81 (2012)
  [arXiv:1112.5451 [hep-ph]].
%
%
\bibitem{Chau:1986jb} 
  L.~L.~Chau and H.~Y.~Cheng,
ÊÊPhys.\ Rev.\ Lett.\  {\bf 56}, 1655 (1986);
%
  L.~L.~Chau and H.~Y.~Cheng,
ÊÊPhys.\ Rev.\ D {\bf 36}, 137 (1987);
%
  L.~L.~Chau and H.~Y.~Cheng,
ÊÊPhys.\ Lett.\ B {\bf 280}, 281 (1992);
%
  J.~L.~Rosner,
  Phys.\ Rev.\ D {\bf 60}, 114026 (1999)
  [hep-ph/9905366];
  M.~Gronau and J.~L.~Rosner,
ÊÊPhys.\ Lett.\ B {\bf 500}, 247 (2001)
ÊÊ[hep-ph/0010237].
%
%
\bibitem{Chiang:2002mr} 
  C.~W.~Chiang, Z.~Luo and J.~L.~Rosner,
ÊÊPhys.\ Rev.\ D {\bf 67}, 014001 (2003)
ÊÊ[hep-ph/0209272];
%
  B.~Bhattacharya and J.~L.~Rosner,
  Phys.\ Rev.\ D {\bf 77}, 114020  (2008)
  [arXiv:0803.2385 [hep-ph]]; 
ÊÊPhys.\ Rev.\ D {\bf 79}, 034016 (2009)
ÊÊ[Erratum-ibid.\ D {\bf 81}, 099903 (2010)]
ÊÊ[arXiv:0812.3167 [hep-ph]];
  Phys.\ Rev.\ D {\bf 81}, 014026 (2010)
  [arXiv:0911.2812 [hep-ph]];
  %
  H.~Y.~Cheng and C.~W.~Chiang,
ÊÊPhys.\ Rev.\ D {\bf 86}, 014014 (2012)
ÊÊ[arXiv:1205.0580 [hep-ph]].
%
%
\bibitem{Buccella:1990sp} 
  F.~Buccella, M.~Forte, G.~Miele and G.~Ricciardi,
  Z.\ Phys.\ C {\bf 48}, 47 (1990);
  F.~Buccella, M.~Lusignoli, G.~Miele and A.~Pugliese,
  Z.\ Phys.\ C {\bf 55}, 243 (1992);
  F.~Buccella, M.~Lusignoli, G.~Miele, A.~Pugliese and P.~Santorelli,
  Phys.\ Rev.\ D {\bf 51}, 3478 (1995)
  [hep-ph/9411286];
  M.~Zhong, Y.~L.~Wu and W.~Y.~Wang,
  Eur.\ Phys.\ J.\ C {\bf 32S1}, 191 (2004);
%
  Y.~L.~Wu, M.~Zhong and Y.~F.~Zhou,
  Eur.\ Phys.\ J.\ C {\bf 42}, 391 (2005)
  [hep-ph/0405080];
%
  D.~N.~Gao,
  Phys.\ Lett.\ B {\bf 645}, 59 (2007)
  [hep-ph/0610389].
%
%
\bibitem{Hiller:2012xm} 
  G.~Hiller, M.~Jung and S.~Schacht,
ÊÊPhys.\ Rev.\ D {\bf 87}, 014024 (2013)
ÊÊ[arXiv:1211.3734 [hep-ph]].
%
%
\bibitem{Grossman:2012ry}
 Y.~Grossman and D.~J.~Robinson,
  JHEP {\bf 1304}, 067 (2013)
  [arXiv:1211.3361 [hep-ph]].
  %
  %
  %
  \bibitem{Gronau:2013xba} 
  M.~Gronau,
  Phys.\ Lett.\ B {\bf 730}, 221 (2014)
  [Addendum-ibid.\ B {\bf 735}, 282 (2014)]
  [arXiv:1311.1434 [hep-ph]].
  \bibitem{Gronau:2014nda} 
  M.~Gronau,
  Phys.\ Rev.\ D {\bf 90}, 117901 (2014)
  [arXiv:1410.7255 [hep-ph]].
  %
  \bibitem{isospinsecond} In Ref. \cite{Gronau:2013xba} it was stated without proof 
  that these isospin breaking terms are first order SU(3) breaking.
%
  \bibitem{Meshkov:1964zz} 
  S.~Meshkov, G.~A.~Snow and G.~B.~Yodh,
  Phys.\ Rev.\ Lett.\  {\bf 13}, 212 (1964);
  H.~J.~Lipkin,
  Phys.\ Rev.\  {\bf 174}, 2151 (1968).
  %
\bibitem{Kingsley} 
R.~L.~Kingsley, S.~B.~Treiman, F.~Wilczek and A.~Zee,
Ref.\,\cite{Kingsley:1975fe}.
\bibitem{Gronau:2000ru} 
  M.~Gronau and J.~L.~Rosner,
 Ref.\,\cite{Chau:1986jb}.
%
\bibitem{Gronau:2012kq}
  M.~Gronau and J.~L.~Rosner,
Phys.\ Rev.\ D {\bf 86}, 114029 (2012)
  [arXiv:1209.1348 [hep-ph]].
\bibitem{Falk:2001hx} 
  A.~F.~Falk, Y.~Grossman, Z.~Ligeti and A.~A.~Petrov,
  Phys.\ Rev.\ D {\bf 65}, 054034 (2002)
  [hep-ph/0110317].  
  %
  \bibitem{Agashe:2014kda} 
  K.~A.~Olive {\it et al.}  [Particle Data Group Collaboration],
  Chin.\ Phys.\ C {\bf 38}, 090001 (2014).
%
  \bibitem{Brod:2011re} 
  See e. g.
  J.~Brod, A.~L.~Kagan and J.~Zupan,
  Phys.\ Rev.\ D {\bf 86}, 014023 (2012)
  [arXiv:1111.5000 [hep-ph]].
  %
  \bibitem{Golden:1989qx} 
  M.~Golden and B.~Grinstein,
  Phys.\ Lett.\ B {\bf 222}, 501 (1989).
 %
%
%
%
\bibitem{Bhattacharya:2012ah} 
  B.~Bhattacharya, M.~Gronau and J.~L.~Rosner,
  Phys.\ Rev.\ D {\bf 85}, 054014 (2012)
  [arXiv:1201.2351 [hep-ph]];
 B.~Bhattacharya, M.~Gronau and J.~L.~Rosner, Proceedings of the Tenth International 
 Conference on Flavor Physics and CP Violation - FPCP2012, May 21-- 25 2012, Hefei, 
 SLAC eConf C120521
 [arxiv:1207.0761[hep-ph]].
%
%
\bibitem{Suzuki:1979uf} 
  M.~Suzuki,
ÊÊPhys.\ Rev.\ Lett.\  {\bf 43}, 818 (1979).
%
\bibitem{Abbott:1979fw}
  L.~F.~Abbott, P.~Sikivie and M.~B.~Wise,
  Phys.\ Rev.\ D {\bf 21}, 768 (1980).
%
\bibitem{Savage}
M. J. Savage, Ref.\,\cite{Savage:1991wu}.
%
%
 \bibitem{Feldmann:2012js} 
 T.~Feldmann, S.~Nandi and A.~Soni,
JHEP {\bf 1206}, 007  (2012) 
  [arXiv:1202.3795 [hep-ph]].
  %
  \bibitem{Brod:2012ud} 
  J.~Brod, Y.~Grossman, A.~L.~Kagan and J.~Zupan,
  JHEP {\bf 1210}, 161 (2012)
  [arXiv:1203.6659 [hep-ph]].
  %
\bibitem{GrosRob}  It was noted in Ref.\,\cite{Grossman:2012ry} that first order U-spin 
breaking cancels in $R_3$.
%
\bibitem{MGJR} It was shown in Ref.\,\cite{Gronau:2012kq} that first order U-spin 
breaking cancels also in $|A(D^0\to K^+K^-)|^2 + |A(D^0 \to \pi^+\pi^-)|^2 
- 2|A(D^0 \to \pi^+K^-)||A(D^0\to K^+\pi^-)|$ contributing to $D^0$--$\bar D^0$ mixing.
 \bibitem{Antonelli:2009ws} 
  M.~Antonelli, D.~M.~Asner, D.~A.~Bauer, T.~G.~Becher, M.~Beneke, A.~J.~Bevan, M.~Blanke and C.~Bloise {\it et al.},
  Phys.\ Rept.\  {\bf 494}, 197 (2010)
  [arXiv:0907.5386 [hep-ph]].
%
 \bibitem{Aaltonen:2013pja} 
  T.~A.~Aaltonen {\it et al.}  [CDF Collaboration],
  Phys.\ Rev.\ Lett.\  {\bf 111}, 231802 (2013)
 [arXiv:1309.4078 [hep-ex]];
  R.~Aaij {\it et al.}  [LHCb Collaboration],
  Phys.\ Rev.\ Lett.\  {\bf 111}, 251801 (2013)
  [arXiv:1309.6534 [hep-ex]];
  B.~R.~Ko {\it et al.}  [Belle Collaboration],
  Phys.\ Rev.\ Lett.\  {\bf 112}, 111801 (2014)
  [arXiv:1401.3402 [hep-ex]]. 
%
 \bibitem{Asner:2003uz} 
  D.~M.~Asner {\it et al.}  [CLEO Collaboration],
  Phys.\ Rev.\ D {\bf 70}, 091101 (2004)
  [hep-ex/0311033];
  B.~Aubert {\it et al.}  [BaBar Collaboration],
  Phys.\ Rev.\ D {\bf 78}, 034023 (2008)
  [arXiv:0804.2089 [hep-ex]].
%
\bibitem{Tian:2005ik} 
  X.~C.~Tian {\it et al.}  [Belle Collaboration],
  Phys.\ Rev.\ Lett.\  {\bf 95}, 231801 (2005)
  [hep-ex/0507071].
  %
  \bibitem{Aubert:2006kt} 
  B.~Aubert {\it et al.}  [BaBar Collaboration],
  Phys.\ Rev.\ Lett.\  {\bf 97}, 221803 (2006)
  [hep-ex/0608006].
  %
\bibitem{Aubert:2008zh} 
  B.~Aubert {\it et al.}  [BaBar Collaboration],
  Phys.\ Rev.\ Lett.\  {\bf 103}, 211801 (2009)
  [arXiv:0807.4544 [hep-ex]].
  %
  \bibitem{GRob} A similar prediction assuming $R''_3=1$ with no uncertainty was 
  obtained in Ref. \cite{Grossman:2012ry}.
  %
  \bibitem{neutrals} See e.g. C. Quigg in Ref.\, \cite{Kingsley:1975fe} and 
  Ref.\, \cite{Gronau:2012kq}. 
  %
  \bibitem{Bigi:1994aw} 
  I.~I.~Y.~Bigi and H.~Yamamoto,
  Phys.\ Lett.\ B {\bf 349}, 363 (1995)
  [hep-ph/9502238].
  %
  \bibitem{eta_mixing} See e.g. F.~G.~Cao,
  Phys.\ Rev.\ D {\bf 85}, 057501 (2012)
  [arXiv:1202.6075 [hep-ph]] and references therein.
  %
    \bibitem{He:2007aj} 
  Q.~He {\it et al.}  [CLEO Collaboration],
  Phys.\ Rev.\ Lett.\  {\bf 100}, 091801 (2008)
  [arXiv:0711.1463 [hep-ex]].
  %
  \bibitem{Rosner:2006bw} 
  J.~L.~Rosner,
  Phys.\ Rev.\ D {\bf 74}, 057502 (2006)
  [hep-ph/0607346].
  %
  \bibitem{Gronau:2014pra} 
  M.~Gronau,
  Phys.\ Lett.\ B {\bf 738}, 136 (2014)
  [arXiv:1407.7374 [hep-ph]].
  
\end{thebibliography}
\end{document}